\documentclass[aps,prd,amsmath,amssymb,footinbib]{revtex4-1}

\usepackage{graphicx}
\usepackage{natbib}

\usepackage{bm}
\usepackage{url}
\usepackage{textcase}
\usepackage{subcaption}
\usepackage{float}
\usepackage{amsthm}
\usepackage{braket}
\usepackage{color}

\newcommand{\be}{\begin{equation}}
\newcommand{\ee}{\end{equation}}

\begin{document}

\title{Circular Orbits Structure and Thin Accretion Disks around Kerr Black Holes with Scalar Hair}

\author{Lucas G. Collodel}
 \email{lucas.gardai-collodel@uni-tuebingen.de}

\affiliation{Theoretical Astrophysics, Eberhard Karls University of T\"ubingen, T\"ubingen 72076, Germany}

\author{Daniela D. Doneva}
\email{daniela.doneva@uni-tuebingen.de}
\affiliation{Theoretical Astrophysics, Eberhard Karls University of T\"ubingen, T\"ubingen 72076, Germany}

\author{Stoytcho S. Yazadjiev}
\email{yazad@phys.uni-sofia.bg}
\affiliation{Theoretical Astrophysics, Eberhard Karls University of T\"ubingen, T\"ubingen 72076, Germany}
\affiliation{Institute of Mathematics and Informatics, 	Bulgarian Academy of Sciences, 	Acad. G. Bonchev St. 8, Sofia 1113, Bulgaria}

\begin{abstract}
In this paper we first investigate the equatorial circular orbit structure of Kerr black holes with scalar hair (KBHsSH) and highlight their most prominent features which are quite distinct from the exterior region of ordinary bald Kerr black holes, i.e. peculiarities that arise from the combined bound system of a hole with an off-center, self-gravitating distribution of scalar matter. Some of these traits are incompatible with the thin disk approach, thus we identify and map out various regions in the parameter space respectively. All the solutions for which the stable circular orbital velocity (and angular momentum) curve is continuous are used for building thin and optically thick disks around them, from which we extract the radiant energy fluxes, luminosities and efficiencies. We compare the results in batches with the same spin parameter $j$ but different normalized charges, and the profiles are richly diverse. Because of the existence of a conserved scalar charge, $Q$, these solutions are non-unique in the $(M, J)$ parameter space. Furthermore, $Q$ cannot be extracted asymptotically from the metric functions. Nevertheless, by constraining the parameters through different observations, the luminosity profile could in turn be used to constrain the Noether charge and characterize the spacetime, should KBHsSH exist.
\end{abstract}

\maketitle

\section{INTRODUCTION}
The last few years have been very exciting for black hole (BH) physics as new ways of observing them became reality. Since 2016, there have been over forty confident BH-BH merging events detected \cite{Abbott:2020gyp}, and more recently the first imaging of a BH contender has been realized \cite{Akiyama:2019cqa}, for the compact object powering M87. Moreover, hundreds of X-ray binaries, many of which with a BH candidate at the center, have been detected thus far thanks to missions like RXTE and Suzaku, and two decades long observations of stars moving at the center of our Milky Way indicate that a massive BH is dwelling there \cite{Ghez:2008ms,Genzel:2010zy}. While electromagnetic observations leave room for BH mimickers as viable alternatives, the wave forms obtained from gravitational wave detectors strongly support the existence of BHs. Despite all of this success, the question of whether general relativity (GR) is the most accurate classical theory of gravity, or whether astrophysically relevant BHs are uniquely described by the Kerr solution is not yet fully answered. 

Half a century ago, a series of theorems laid the ground for the \emph{Kerr hypothesis} \cite{PhysRev.164.1776, PhysRevLett.26.331,PhysRevLett.34.905}, for according to these \emph{no-hair theorems} the only stationary, axisymmetric, asymptotically flat, regular outside of the horizon solution to four-dimensional GR when the matter fields feature the same isometries of the spacetime is the Kerr BH. Notwithstanding their significance, there are many ways with which to circumvent them and discover different solutions. Still in four dimensions, hairy BHs have been described in different theories of gravity, such as Einstein-Yang-Mills \cite{Volkov:1990sva,doi:10.1063/1.528773,PhysRevLett.64.2844,BREITENLOHNER1992357,PhysRevD.57.6138,PhysRevLett.86.3704,KLEIHAUS2004294}, scalar-tensor \cite{Bocharova:1970skc,BEKENSTEIN1974535,KLEIHAUS2015406,PhysRevD.102.084032} and Gauss-Bonnet theories \cite{PhysRevD.54.5049,PhysRevLett.106.151104,PhysRevD.93.044047,PhysRevLett.120.131104,PhysRevLett.120.131103,PhysRevLett.120.131102,PhysRevLett.123.011101,Collodel_2020,Berti:2020kgk,Herdeiro:2020wei}. Remarkably, by dropping the assumption that the matter fields must be stationary and axisymmetric, Herdeiro and Radu found solutions in the context of GR where BHs have hair \cite{PhysRevLett.112.221101,Herdeiro_2015}, by minimally coupling to gravity a complex scalar field which depends on time and on the axial coordinate while its energy-momentum tensor still possesses the respective isometries, see \cite{PhysRevD.92.084059,doi:10.1142/S0218271816410145,Herdeiro_2016,DELGADO2016234,BRIHAYE2016279} for generalizations. These are known as scalarized Kerr black holes (KBHsSH) and they are the object of study of this paper. In their domain of existence they connect Kerr BHs (that is with no hair) with pure solitonic solutions, also known as boson stars (BS), which are regular everywhere and feature no horizons. In this sense, one can think of the KBHsSH indeed as a combined system of a BS with a horizon at its center, and therefore it shares traits of both objects. 

BSs are very peculiar stars that first appeared in the literature in the late sixties \cite{PhysRev.187.1767,PhysRev.172.1331}, and they are the realization of a complex scalar field bound by its self-gravity. These objects are not enveloped by a defined surface where the pressure becomes zero. Instead, the field extends all the way to infinity with exponential decay. Because they only interact gravitationally with other matter fields, particles can freely move in their interior, where the curvature is large enough for them to fall under the category of compact objects \cite{PhysRevD.85.024045}. Because there is no degenerate pressure for bosonic fields, the maximum mass BSs can achieve depends solely on their self-interaction potential. Similarly their size spans from atomic \cite{PhysRevD.35.3640} to galactic scales where they can in principle mimic BHs in active galactic nuclei (AGN) \cite{Vincent_2016}. When rotating, BSs have a quantized angular momentum proportional to their scalar charge \cite{Schunck1996}, and hence cannot spin slowly in a perturbative sense. Moreover, the topology of the scalar field's profile changes upon rotation, and it distributes itself on a torus. Thereafter, rotating BSs have been extensively studied, along with their domain of existence, geometrical properties, stability and formation \cite{PhysRevD.72.064002,PhysRevD.77.064025,PhysRevD.96.084066,PhysRevD.99.104076,PhysRevLett.123.221101}. A necessary condition for KBHsSH is that the Lie derivative of the scalar field with respect to the Killing vector at the horizon disappears. In other words, the angular velocity of the horizon must be the same as the phase velocity of the scalar field, and therefore this hair is \emph{synchronized}, and there are no static KBHsSH. Geodesics around BSs have been reported in \cite{PhysRevD.90.024068,Meliani_2015,Grould:2017rzz}, and are quite special due to the off-center energy density distribution for the spinning case and the test particles not being restricted to an exterior region. In particular, solutions without ergoregions feature a static ring in the equatorial plane where freely falling matter remains at rest with respect to a zero angular momentum observer at infinity \cite{PhysRevLett.120.201103}.

The theory of accretion disks is diverse with many different approaches, see \cite{Abramowicz:2011xu} for a review. The thin disk model was introduced in \cite{Shakura:1972te} with the so called $\alpha$-disk and had been shortly after given a relativistic description \cite{Novikov:1973kta,Page:1974he}. It is built upon several simplifying, reasonable assumptions with which quantities of interest such as the radiant energy flux and the luminosity of the disk are given in integral form. Specifically, the disk is thin enough so that any function is evaluated on the equatorial plane; it is optically thick and radiation flowing within the disk is negligible; it is in thermal equilibrium and radiates as a black body; and particles move on almost circular geodesics until the innermost stable circular orbit (ISCO) where the plunging happens, but the radial velocity (and thus the accretion rate) is an ad-hoc factor. Thin disk model has been applied to BHs in various theories such as $f(R)$ \cite{PhysRevD.78.024043,Perez:2012bx} (also for neuton stars \cite{Staykov_2016}), Chern-Simons \cite{Harko_2010}, scalar-tensor-vector \cite{PhysRevD.95.104047} and Einstein-Maxwell-Dilaton \cite{Heydari-fard:2020syf}, as well as for BSs in GR \cite{TORRES2002377,PhysRevD.73.021501}. Accretion disk theory is of great importance in astronomy as it serves as a template for parameter estimation via data fitting. For instance, thin disk models are used in determining the angular momentum of a BH, either via the continuum-fitting method, or via X-ray reflection spectroscopy \cite{Zhang_1997,McClintock_2011,McClintock:2013vwa,Brenneman_2006,Reynolds:2013qqa}. It has recently been pointed out that there is a difference in this estimation when compared with results produced with slim disk (finite thickness) models, but small compared to the present observational uncertainties \cite{10.1093/mnras/staa1591}. In the context of BH mergers, the spin of the final object can be estimated from the ISCO properties of the original holes \cite{PhysRevD.77.026004,PhysRevD.96.044031}. 
The extraction of K$\alpha$ iron line information also relies on thin disk models, and provides a framework with which to distinguish between different astrophysical objects \cite{Johannsen:2012ng,PhysRevD.87.023007,PhysRevD.88.064022,Jiang_2015,Cao_2016,Bambi_2017}.  In \cite{Ni_2016}, the authors explored the K$\alpha$ iron line of a sample of three quite different solutions of KBHsSH to see if they could, as a template, fit the fabricated data produced by a Kerr BH and found that two of them performed well enough to be dismissed. Thick tori (a generalization of the Polish Doughnut) have also been constructed for KBHsSH \cite{PhysRevD.99.043002} and BSs \cite{Meliani_2015,Teodoro:2020kok}, which can be used as initial data for dynamical simulations as well as for modeling the imaging of a dark compact object.
Dynamical simulations of accreting disks and disruption events around BSs  in comparison to BHS have been considered in \cite{10.1093/mnras/staa1878,Teodoro:2020gps}, while the possibility of constraining the gravitational theory and object via our current imaging capacities is the subject of study in \cite{Mizuno:2018lxz}.

The paper is organized as follows. In section \ref{sec:theory} we revisit the general theory behind KBHsSH, circular orbits in the equatorial plane and thin accretion disks. The structure of circular orbits in KBHsSH spacetimes is reported in \ref{sec:metricisco}, where the peculiarities arising from the system combination of a Kerr BH with a BS become apparent. The main results of applying thin disks on KBHsSH are given in section \ref{sec:results} which is followed by our conclusions in section \ref{sec:conclusions}. Unless otherwise specified, we employ units where $c=G=\hbar=1$.

\section{THEORY}
\label{sec:theory}

\subsection{Kerr Black Holes with Scalar Hair}
The spacetimes we investigate are described from first principles by minimally coupling a complex scalar field to gravity via the action
\begin{equation}
\label{action}
S=\int \left[\frac{R}{2}-g^{\mu\nu}\partial_\mu\Phi^*\partial_\nu\Phi-2U(\Phi)\right]\sqrt{-g}d^4x, 
\end{equation}
where $R$ is the Ricci curvature, $g$ is the metric determinant and $\Phi$ is a complex scalar field whose mass and self interaction are determined by the potential $U$.
Therefore, the scalar field acts solely as a source field of non-baryonic matter and the underlying theory is still general relativity. The Einstein field equations are obtained as usual by varying the action with respect to $g^{\mu\nu}$
\begin{equation}
\label{efe}
R_{\mu\nu}-\frac{1}{2}Rg_{\mu\nu}=-g_{\mu\nu}\left(\partial^\sigma\Phi^*\partial_\sigma\Phi+2U\right)+2\partial_\mu\Phi^*\partial_\nu\Phi,
\end{equation}
where the right hand side is simply the energy-momentum tensor of the scalar matter. Varying the action with respect to $\Phi$ and $\Phi^*$ yields a pair of Klein-Gordon equations, 
\begin{equation}
\label{kge}
\left(\Box+2\frac{\partial U}{\partial|\Phi|^2}\right)\Phi^*=0, \qquad \left(\Box+2\frac{\partial U}{\partial|\Phi|^2}\right)\Phi=0,
\end{equation}
to be solved simultaneously with the field equations. The system is invariant under $U(1)$ transformations, $\Phi\rightarrow\Phi e^{i\alpha}$, and therefore possesses a conserved Noether current,
\begin{equation}
\label{ncurrent}
j^\mu=-i(\Phi^*\partial^\mu\Phi-\Phi\partial^\mu\Phi^*),
\end{equation}
which, when projected over the future timelike normal direction $n_\mu$ and integrated over the spacelike three-volume bounded by the horizon surface and infinity $\Sigma\backslash\mathcal{H}$, gives a conserved charge,
\begin{equation}
\label{ncharge}
Q=\int_{\Sigma\backslash\mathcal{H}}j^\mu n_\mu dV,
\end{equation}
where $dV$ is the volume element.

We are interested in stationary and axisymmetric solutions, and choose a metric form in adapted spherical coordinates $\{t, r, \theta, \varphi\}$ given by
\be
\label{ds1}
ds^2=-\mathcal{N}e^{2F_0}dt^2+e^{2F_1}\left(\frac{dr^2}{\mathcal{N}}+d\theta^2\right)+e^{2F_2}r^2\sin^2\theta\left(d\varphi-\omega dt\right)^2,
\ee
for which $\{F_0, F_1, F_2, \omega\}$ is the set of functions dependent on both $r$ and $\theta$ that we solve for, and $\mathcal{N}\equiv 1-r_H/r$, where $r_H$ is the coordinate size of the horizon. In this coordinates, the Killing vector fields are $\xi^\mu=\vec{\partial}_t$ and $\chi^\mu=\vec{\partial}_\varphi$ corresponding respectively to stationarity and axisymmetry. The chosen aesthetics of the line element is not of importance for the work developed here, but the choice of coordinates is. Therefore, we will refer to the metric functions written in the more general way
\be
\label{dst2}
ds^2=g_{tt}dt^2+g_{rr}dr^2+g_{\theta\theta}d\theta^2+2g_{t\varphi}dtd\varphi+g_{\varphi\varphi}d\varphi^2.
\ee

The underlying mechanism to give rise to scalarization is a superradiant instability that only occurs for rotating systems. To level the scalar field with a fluid it is necessary to assign to it a four velocity, and therefore it must depend on all four spacetime coordinates. Stationarity and axisymmetry, however, require that the field's dependence on $t$ and $\varphi$ be given in an explicit way such that
\begin{equation}
\Phi=\phi(r,\theta)e^{i(\omega_st+m\varphi)},
\end{equation}
where $\omega_s$ is its natural frequency and $m$ is the winding number, belonging to the integer set. A further criteria for scalarization is obtained by evolving the linearized version of eq. (\ref{kge}) on a fixed Kerr background and observing where the threshold between stable and unstable modes lies, i.e. where $\omega_s$ is real, which is demanded for stationary solutions. This occurs only when the scalar field's angular velocity matches that of the horizon, explicitly 
\begin{equation}
\frac{\partial_t\Phi}{\partial_\varphi\Phi}=-\frac{g_{t\varphi}}{g_{\varphi\varphi}}\rightarrow \frac{\omega_s}{m}=\omega_H.
\end{equation}

As for the potential, we consider non-interacting fields 
\begin{equation}
\label{spot}
U=\frac{1}{2}\mu^2\vert\Phi\vert^2,
\end{equation}
where $\mu$ is the mass of the boson, that scales the equations through the transformations
\begin{equation}
\label{mus}
r\rightarrow r\mu, \qquad \omega_s\rightarrow \omega_s/\mu, \qquad M\rightarrow M\mu,
\end{equation}
which are implied in the rest of the text.

The ADM mass and angular momentum can be extracted asymptotically from the metric functions' leading terms at infinity,
\begin{equation}
\label{asympmj}
M=\frac{1}{2}\lim_{r\rightarrow\infty}r^2\partial_rg_{tt}, \qquad J=-\frac{1}{2}\lim_{r\rightarrow\infty}r^2g_{t\varphi},
\end{equation}
and can also be calculated through the Komar integrals defined with the spacetime Killing vector fields, which after explicitly breaking down to the contributions given by the hole itself ($M_H, J_H$) and the scalar hair ($M_\phi, J_\phi$) give
\begin{equation}
\label{komar}
M=\underbrace{2\int_\mathcal{H} n^\mu\sigma^\nu\nabla_\mu\xi_\nu dA_H}_{M_H} + \underbrace{2\int_{\Sigma\backslash\mathcal{H}} n^\mu\nabla_\nu\nabla_\mu\xi^\nu dV}_{M_\phi},\qquad J=\underbrace{-\int_\mathcal{H} n^\mu\sigma^\nu\nabla_\mu\eta_\nu dA_H}_{J_H} \underbrace{-\int_{\Sigma\backslash\mathcal{H}} n^\mu\nabla_\nu\nabla_\mu\eta^\nu dV}_{J_\phi},
\end{equation} 
where $\mathcal{H}$ is the horizon surface, $dA_H$ its area element and $\sigma^\mu$ a spacelike vector perpendicular to it such that $\sigma^\mu\sigma_\mu=1$. 

The angular momentum stored in the hair can be expressed as an integer multiple of the total charge, $J_\phi=mQ$. It is useful then to define a normalized charge as a measure of how hairy a particular solution is $q\equiv J_\phi/mJ_H$. The higher the winding number $m$ is, the more angularly excited the solutions become. Hence, we restrict our analysis to, the case $m=1$.

\subsection{Circular Orbits on the Equatorial Plane}

In the thin disk approach, each volume element of the fluid moves freely in circular orbit on the equatorial plane, while the vertical profile of the disk and the accretion process itself are taken in a phenomenological fashion, e.g. see section \ref{sec:TD}. Hence we revisit next the equations describing this kind of geodesic motion and how the physical quantities are related. On the equatorial plane $\theta=\pi/2$, $d\theta=0$ and the line element reads simply
\begin{equation}
ds^2_{\pi/2}=g_{tt}(r,\pi/2)dt^2+g_{rr}(r,\pi/2)dr^2+2g_{t\varphi}(r,\pi/2)dtd\varphi+g_{t\varphi}(r,\pi/2)d\varphi^2.
\end{equation}

Each constituent of the fluid is a timelike particle in circular motion whose four-velocity is $u^\mu=u^t(\xi^\mu+\Omega\chi^\mu)$, where $\Omega:=u^\varphi/u^t$ is the orbital velocity. Due to stationarity and axisymmetry, the particle's normalized energy $E=-p_t$ and angular momentum $L=p_\varphi$ are conserved. Therefore, we have 
\begin{equation}
\label{4vel}
u^t = -\frac{g_{t\varphi}L+g_{\varphi\varphi}E}{g_{t\varphi}^2-g_{tt}g_{\varphi\varphi}}; \qquad u^\varphi = \frac{g_{tt}L+g_{t\varphi}E}{g_{t\varphi}^2-g_{tt}g_{\varphi\varphi}}.
\end{equation}

The four velocity norm $u^\mu u_\mu=-1$ can be used to derive the equation of motion for the radial coordinate, and it gives
\begin{equation}
\label{eqrd}
V_{eff}\equiv g_{rr}\dot{r}^2=\frac{g_{\varphi\varphi}}{g_{t\varphi}^2-g_{tt}g_{\varphi\varphi}}\left(E-V_+\right)\left(E-V_-\right),
\end{equation}
where we define the effective potential $V_{eff}$ in terms of the potentials $V_\pm$ 
\begin{equation}
\label{vpm}
V_\pm\equiv L\frac{g_{t\varphi}}{g_{\phi\phi}}\pm\frac{\sqrt{L^2\left(g_{t\varphi}^2-g_{tt}g_{\varphi\varphi}\right)}}{g_{\phi\phi}}.
\end{equation}

This differential equation can be solved with a suitable initial condition for pairs of energy and angular momentum to describe geodesics on the equatorial plane. Circular orbits, characterized by a fixed radius, demand that both $V_{eff}$ and $\partial_rV_{eff}$ be zero. Thus, with these two constraints there is a fixed pair of energy and angular momentum for each circular orbit radius. Using the angular velocity as a constraint, we can write
\begin{equation}
\label{eqen}
E=-\frac{g_{tt}+g_{t\varphi}\Omega}{\sqrt{-g_{tt}-2g_{t\varphi}\Omega-g_{\varphi\varphi}\Omega^2}},
\end{equation}
\begin{equation}
\label{eqan}
L=\frac{g_{t\varphi}+g_{\varphi\varphi}\Omega}{\sqrt{-g_{tt}-2g_{t\varphi}\Omega-g_{\varphi\varphi}\Omega^2}},
\end{equation}
\begin{equation}
\label{eqom}
\Omega=\frac{-\partial_rg_{t\varphi}\pm\sqrt{(\partial_rg_{t\varphi})^2-\partial_rg_{tt}\partial_rg_{\varphi\varphi}}}{\partial_rg_{\varphi\varphi}},
\end{equation}

The ISCO dwells at at smallest radius for which the orbit is marginally stable, meaning $\partial^2_rV_{eff}=0$. Thus, the equation
\begin{equation}
\label{vrr}
E^2\partial_r^2g_{\varphi\varphi}+2EL\partial^2_rg_{t\varphi}+L^2\partial_r^2g_{tt}-\partial^2_r\left(g_{t\varphi}^2-g_{tt}g_{\varphi\varphi}\right)=0,
\end{equation}
together with the aforementioned constraints, is satisfied at $r=r_{ISCO}$.

\subsection{Thin Disks} 
\label{sec:TD}
The thin disk approach was developed in relativistic form in the early seventies \cite{Novikov:1973kta,Page:1974he}. The equation governing the phenomenon are derived from a series of simplifying, yet fairly reasonable assumptions. For the characteristic time scale of the physical process, the background spacetime is unchanged, being stationary, axisymmetric, asymptotically flat and even with respect to reflection onto the equatorial plane. The disk has negligible self-gravity, its central plane lies precisely in the equatorial plane and it is thin, i.e. the height of the accreting disk is negligible compared to its horizontal extension at any given radius, $h\ll r$. Furthermore, radiation is only emitted perpendicularly to the disk's plane, namely it is optically thick. The particles that constitute the fluid are in Keplerian motion and the infalling matter is therefore treated phenomenologically with a mass accretion rate $\dot{M}$ and four-velocity radial component $u^r$ placed in \emph{ad-hoc} manner. The inner edge of the disk is given by the radius of the ISCO and we impose no restriction on the outer boundary other than convergence for the quantities of interest.  In this context, the system is in a steady state, hence $\dot{M}$ is independent of the radial coordinate. The radiation energy $F$ emitted by the disk's surface is given by 

\begin{equation}
\label{eqradflux}
F(r) = -\frac{c^2\dot{M}}{4\pi\sqrt{-g/g_{\theta\theta}}}\frac{\partial_r\Omega}{(E-\Omega L)^2}\int_{r_{ISCO}}^r (E-\Omega L)\partial_rLd,
\end{equation}
and some of the steps necessary for deriving it are recalled in the Appendix \ref{app:refsc}. Since the system is in a steady state it must be in thermodynamical equilibrium emitting as a black body whose temperature profile obeys the law $F=\sigma T^4$, where $\sigma$ is the Stefan-Boltzmann constant. Because for KBHs the flux scales with the BH's mass and accretion rate, we define a normalized flux $\tilde{F}$ to better compare profiles yielded from spacetimes of different spin parameters via
\be
F=\frac{c^2\dot{M}}{M_g^2}\tilde{F},
\ee
where $M_g=GM/c^2$.

The amount of gravitational energy converted into radiation as a particle falls down from infinity all the way to the black hole can be quantified in terms of an efficiency of the central body in doing so. If all photons can escape to infinity, the efficiency can be measured as the difference between the normalized energy measured at infinity and at the ISCO,

\begin{equation} 
\label{eqeff}
\epsilon=1-E_{ISCO}.
\end{equation}

The efficiency for the Schwarzschild BH is just below $6\%$, while for Kerr BHs it increases with the spin parameter reaching $42\%$ in the extremal case. The observed luminosity at a distance $d$ from the source is
\begin{equation} 
\label{eqlum}
\mathcal{L}(\nu) = 4\pi d^2 I(\nu) = \frac{8 h}{\pi c^2}\cos\gamma\int_{r_{ISCO}}^\infty \int_0^{2\pi}\frac{\nu_e^3rdr}{\exp\left(h\nu_e/k_BT\right)-1},
\end{equation}
where $I$ is the Planck's distribution function, $h$ is the Planck constant, $k_B$ is the Boltzmann constant, $\gamma$ is the disk's inclination and $\nu_e$ is the emitted frequency that gets redshifted on its way to the observer, who measures $\nu=\nu_e/(1+z)$, where
\begin{equation}
\label{redshift}
1+z = \frac{1+\Omega r\cos\gamma\sin\varphi}{\sqrt{-g_{tt}-2\Omega g_{t\varphi}-\Omega^2g_{\varphi\varphi}}}.
\end{equation}
Hence we neglect any bending of light and assume the observer is at the asymptotic flat region.

\section{Equatorial Metric Structure and Special Features}
\label{sec:metricisco}
\subsection{Domain of Existence}

On the equatorial plane of most of the studied stationary spacetimes, specially in general relativity, and constraining the analysis to the regions exterior to a horizon (if there is any), the three metric functions that appear in these equations are monotonic, with the clear exception for manifolds with naked singularities. As a consequence, one might intuitively and naively 
take for granted that
\begin{itemize}
\item For all $r>r_{ISCO}$ there exists a circular orbit and it is stable.
\item For all circular orbits, $\Omega_+ >0$ while $\Omega_-<0$ and, therefore, they correspond to the angular velocities of co-rotating and counter-rotating orbits, respectively.
\end{itemize}

Complex scalar fields are angularly excited upon rotation, in a very similar fashion as the hydrogen atom is with increasing magnetic quantum number from zero to one. Thus, the field distributes itself in a torus, with its energy density behaving the same way. This off-center energy configuration causes the spacetime to warp in unusual ways and, in the presence of a black hole at the center, the metric functions might feature different local maxima and minima. The listed items above then fail to be true for a set of solutions and other peculiar features arise, which we shall discuss in detail below. They include, for example, regions where for a fixed radius there are two prograde orbits with different angular velocities and regions where no (inertial) circular orbits are possible at all.

\begin{figure}[h]
\includegraphics[width=0.6\textwidth]{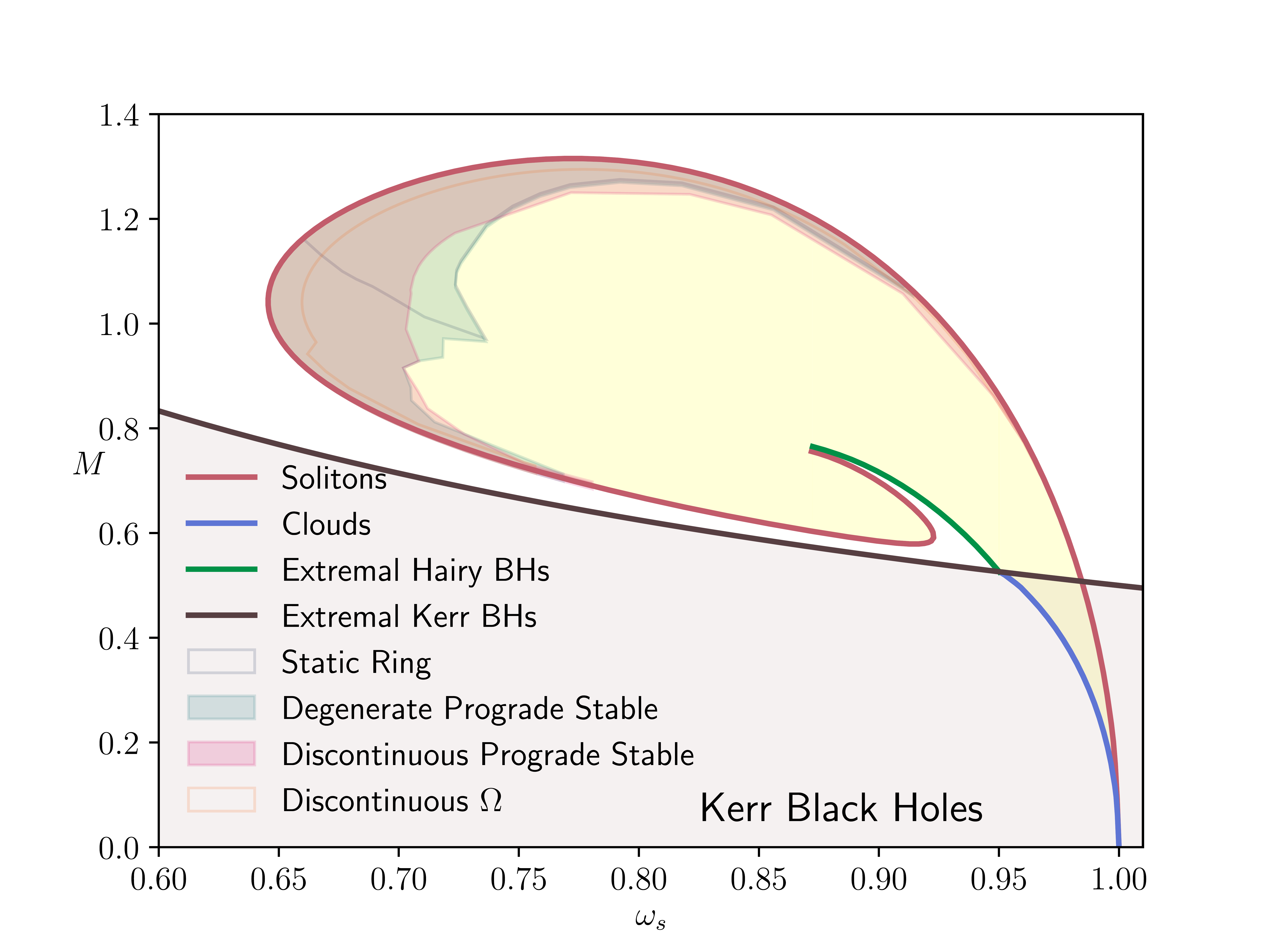}
\caption{Domain of existence for KBHsSH.}
\label{domainskbh}
\end{figure}

The existence domain of KBHsSH is displayed in Fig. \ref{domainskbh} in a $M$ vs. $\omega_s$ diagram. The region is bounded by three qualitatively different sets of solutions. The blue curve corresponds to the clouds, namely Kerr BHs with a marginally bound scalar cloud beyond which the instabilities grow and stationary scalarized solutions appear in the nonlinear regime. On this curve the scalar cloud is not backreacting and therefore the charge is zero, $q=0$. The red curve features no horizon, i.e. it is the set of pure solitonic solutions where $q=1$. Finally, the green curve represents the extremal hairy solutions and one expects it to join the solitonic curve after swirling around each other, in a point near the region in the graph where both stop \cite{PhysRevLett.112.221101,Herdeiro_2015,PhysRevD.102.084032}. Accessing it numerically is notoriously difficult, though. Kerr BHs exist from the solid black curve (extremal holes) below, where the parameters obey
\be
\label{masskerr}
M-\frac{j}{2\omega_s(\sqrt{1-j^2}+1)}=0, \qquad j\equiv \frac{J}{M^2}.
\ee

Within the scalarized solutions there are four overlapping shaded regions where circular orbits on the equatorial plane show unusual features. The greyish area encloses the solutions for which $g_{tt}$ has a local maximum outside an ergoregion, and hence there is a ring of points where $\Omega_-=0$ called the \emph{static ring}. In the greenish region, which overtakes the whole static ring area, lie the solutions which contain an interval of $r$ with degenerate prograde stable orbits, i.e. black holes that at certain distances of the horizon have $\Omega_->0$ and both orbits are stable. The next shaded region in pink features solutions where not all co-rotating circular orbits for $r>r_{ISCO}$ are stable. The last particular set of solutions, in orange, consists of those for which, in certain places (again for $r>r_{ISCO}$), an inertial circular trajectory is not possible. These traits can be readily understood from the equatorial profile of the metric functions, as analyzed below.

\subsection{Shaded Areas in Detail}
\begin{figure}[h]
\includegraphics[width=0.6\textwidth]{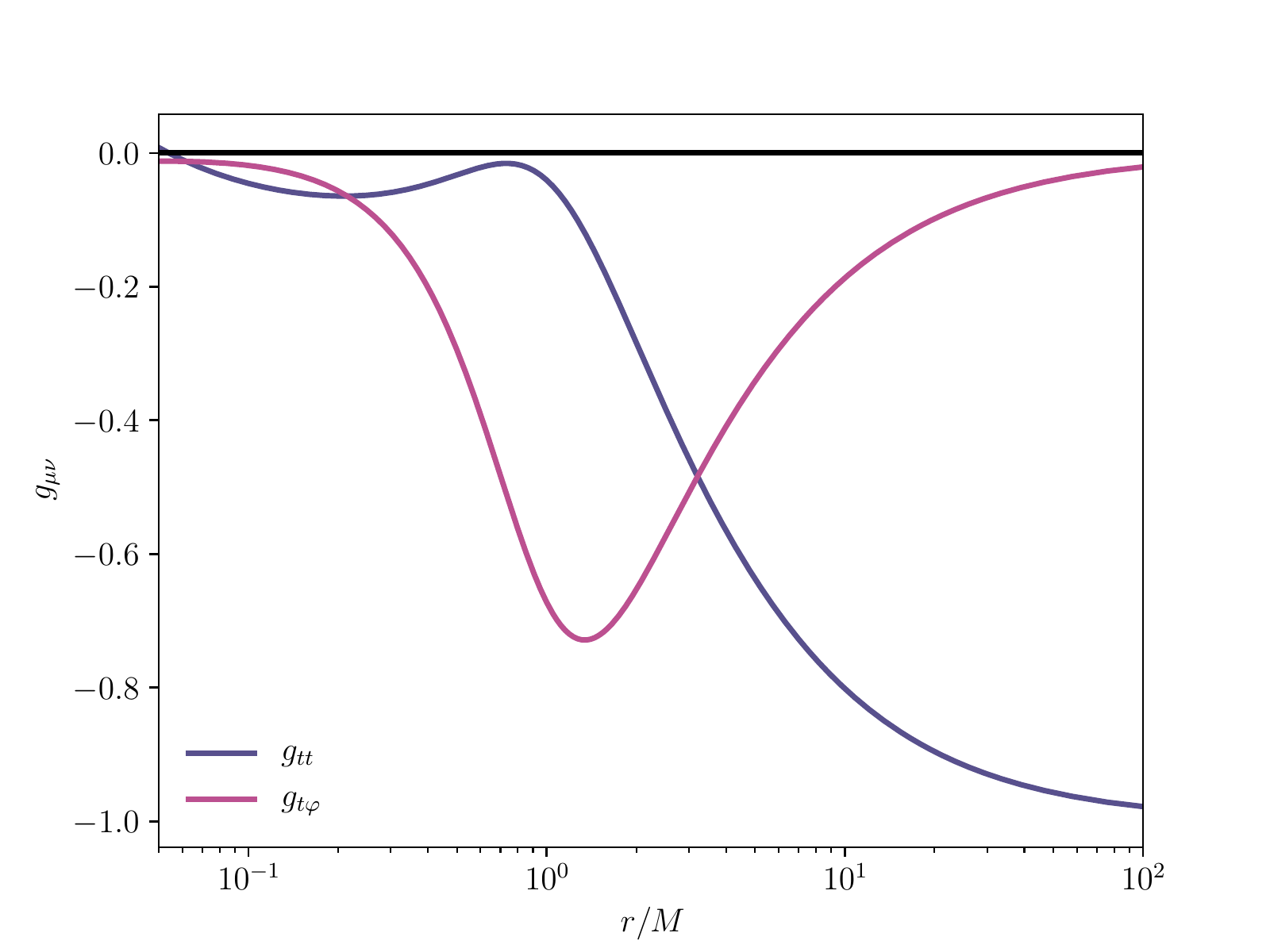}
\caption{Metric components $g_{tt}$ and $g_{t\varphi}$ as functions of the normalized radial coordinate $r/M$ for a solution in the gray hatched region.}
\label{gmex}
\end{figure}

\subsubsection{Static Ring}

On the equatorial plane, if $g_{tt}$ has a local extreme in a region where it is negative (with the metric signature here adopted), then it is clear from eq. (\ref{eqom}) that $\Omega_-=0$ at these points. We remark that if $g_{tt}>0$ this is an ergoregion and static orbits are then only possible for spacelike particles. In Fig. \ref{gmex} we depict the profiles of $g_{tt}$ and $g_{t\varphi}$ for a solution with such feature. In this spacetimes there is both a local minimum and maximum for $g_{tt}$ and it is possible for a test particle to remain at rest in these regions with respect to a static observer at infinity.

\subsubsection{Degenerate Prograde Stable Orbits}

\begin{figure}[h]
\includegraphics[width=0.6\textwidth]{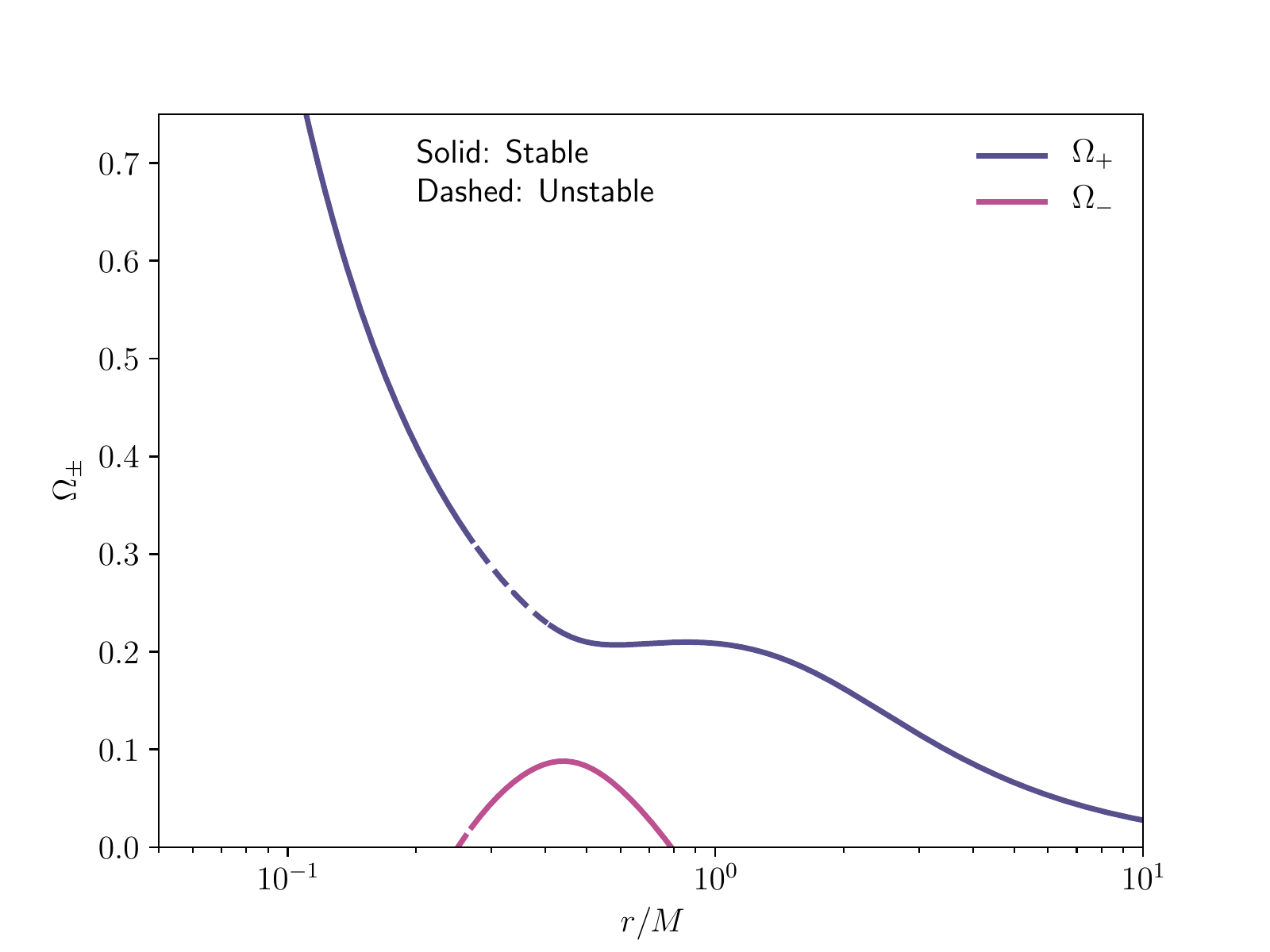}
\caption{Degenerate prograde stable orbits ($\Omega_->0$) and stability discontinuity illustrated. }
\label{gdsd}
\end{figure}

As discussed above, there are certain KBHsSH for which on the equatorial plane there exists a region with no retrogade orbits, i.e. $\Omega_->0$. According to eq. (\ref{eqom}) if the slope of $g_{tt}$ is positive then the quantity in the square root will always be less than $\partial_rg_{tt}$, since the slope of $g_{\varphi\varphi}$ (which is monotonic) is always positive. Observing $g_{tt}$ in Fig. \ref{gmex}, its slope is positive between the two local extreme where the static rings live. Furthermore, it is possible that in some interval of the radial coordinate both prograde orbits are stable. Therefore, there are \emph{degenerate prograde stable orbits}, a feature worth discriminating in the parameter space of Fig. \ref{domainskbh}. To better illustrate this behavior, we provide an example in Fig. \ref{gdsd} where both $\Omega_\pm$ are plotted (only the positive values) and the stable regions are displayed. In this particular case, most part of the radial interval with no retrograde orbits is characterized by stable orbits for both  $\Omega_\pm$.

\subsubsection{Discontinuous Prograde Stable Orbits}

This region contains solutions for which prograde orbits given by $\Omega_+$ suffer from stability discontinuities and is also featured in the example of Fig. \ref{gdsd}. 
Specifically, not all orbits are stable for $r>r_{ISCO}$, but instead there is a region starting and ending with marginally stable orbits where all orbits are unstable.

\subsubsection{Discontinuous Existence of Circular Orbits}

\begin{figure}[h]
\includegraphics[width=0.6\textwidth]{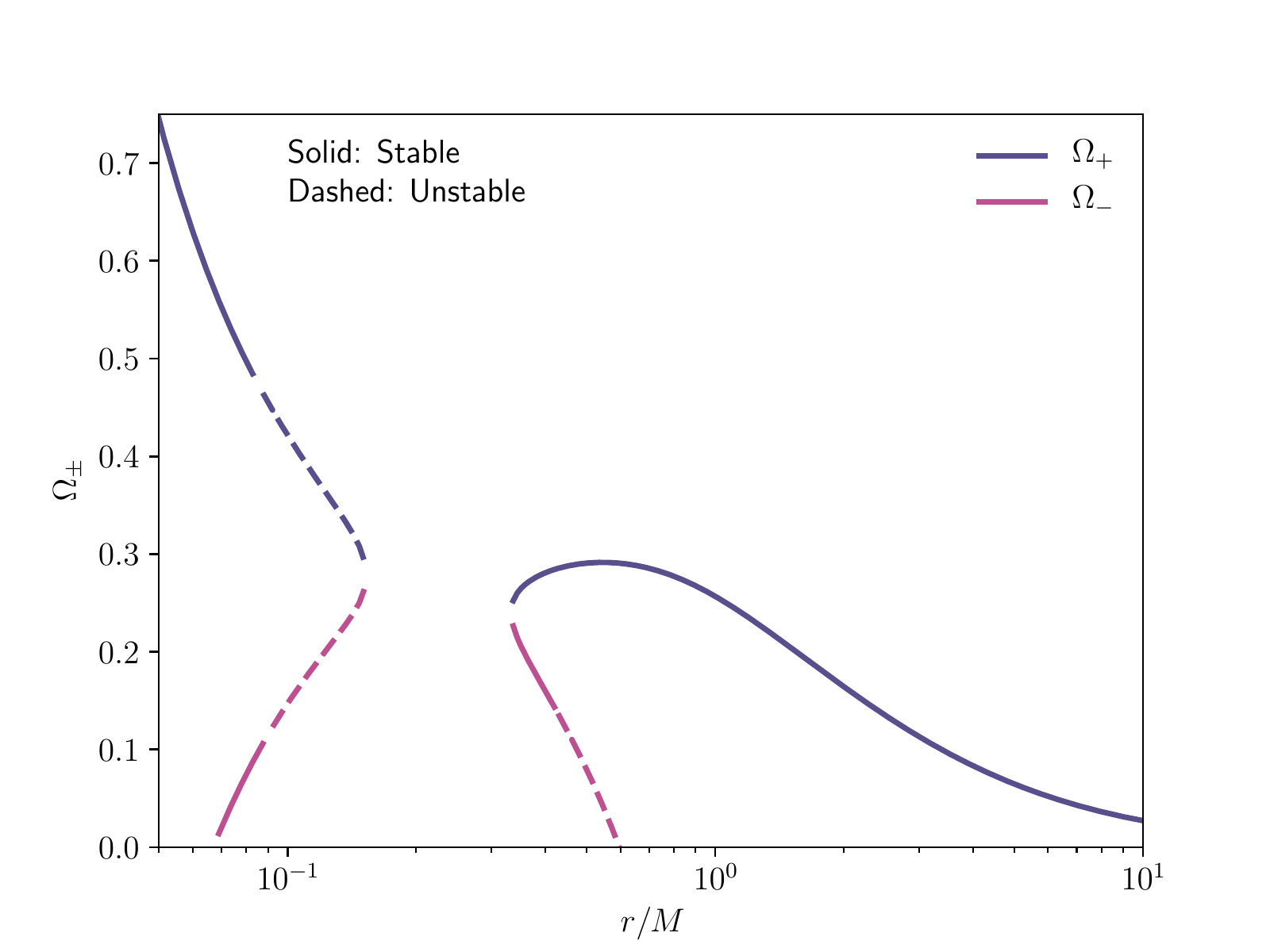}
\caption{Inertial circular orbits cease to exist whenever $\partial_rg_{tt}\partial_rg_{\varphi\varphi}>(\partial_rg_{t\varphi})^2$. }
\label{discomega}
\end{figure}

The three particular cases investigated above arise from the fact that $g_{tt}$ is not monotonic as it usually happens in the exterior region of BHs, namely it can interchange between negative and positive slopes. Things can get even more interesting when the slope is sufficiently positive such that the square root of eq. (\ref{eqom}) becomes zero. This equality occurs when the dragging of the spacetime hits a local extreme and at this point $\Omega_\pm=\omega$ and locally $g_{t\varphi}\propto g_{\varphi\varphi}$. An inertial observer which rotates along with the dragging of the spacetime must have zero angular momentum ($L=0$) an therefore is called a zero angular momentum observer (ZAMO). Usually, ZAMOs have nonzero radial velocity and are therefore not in circular motion. In constrast, as can be seen from equations (\ref{vpm}) and (\ref{eqrd}), in order for them to be in a circular orbit their energy must be zero. One should also note that $\omega$ is the minimum and maximum value that $\Omega_+$ and $\Omega_-$ can achieve, respectively. After this point $\Omega_\pm$ becomes complex and circular orbits are no longer possible for inertial particles until the slope of $g_{tt}$ becomes small enough for yet another point of circular orbiting ZAMOs to appear. Again, this phenomenon is due to the off-center energy density distribution of the scalar hair. There is a gravitational potential well for each the hole and the hair, and angular momentum only contributes to a centrifugal force that prevents the particle to maintain itself at a fixed radius without accelerating. Fig. \ref{discomega} depicts such interrupted existence of circular orbits.

\section{Results}
\label{sec:results}
The thin disk approach is employed for the solutions lying outside the shaded areas displayed in Fig \ref{domainskbh}, where the orbital velocity and angular momentum of prograde moving test particles are continuous. Solutions inside the shaded area require a more careful investigation which might only be possible through dynamical simulations in some cases. For instance, if $\Omega$ is discontinuous, the flux is badly defined. Moreover, it would not be reasonable to assume that accreting particles would keep on near Keplerian orbits in regions of instability or of degenerate prograde stable orbits. One could, however, assume that the ISCO dwells at the radius of the \emph{last} marginally stable circular orbit, after which all orbits are stable and not degenerate. But such approach ignores all the dynamical processes occurring between the horizon and this newly defined ISCO radius, which could host interesting and rich physics whose outcome could possibly be strong enough to shadow the observables extracted from this outer disc. Therefore, a proper investigation on these peculiar solutions will be carried out in future work.

\subsection{Radiant Energy Flux}

\begin{figure}[h]
\includegraphics[width=0.6\textwidth]{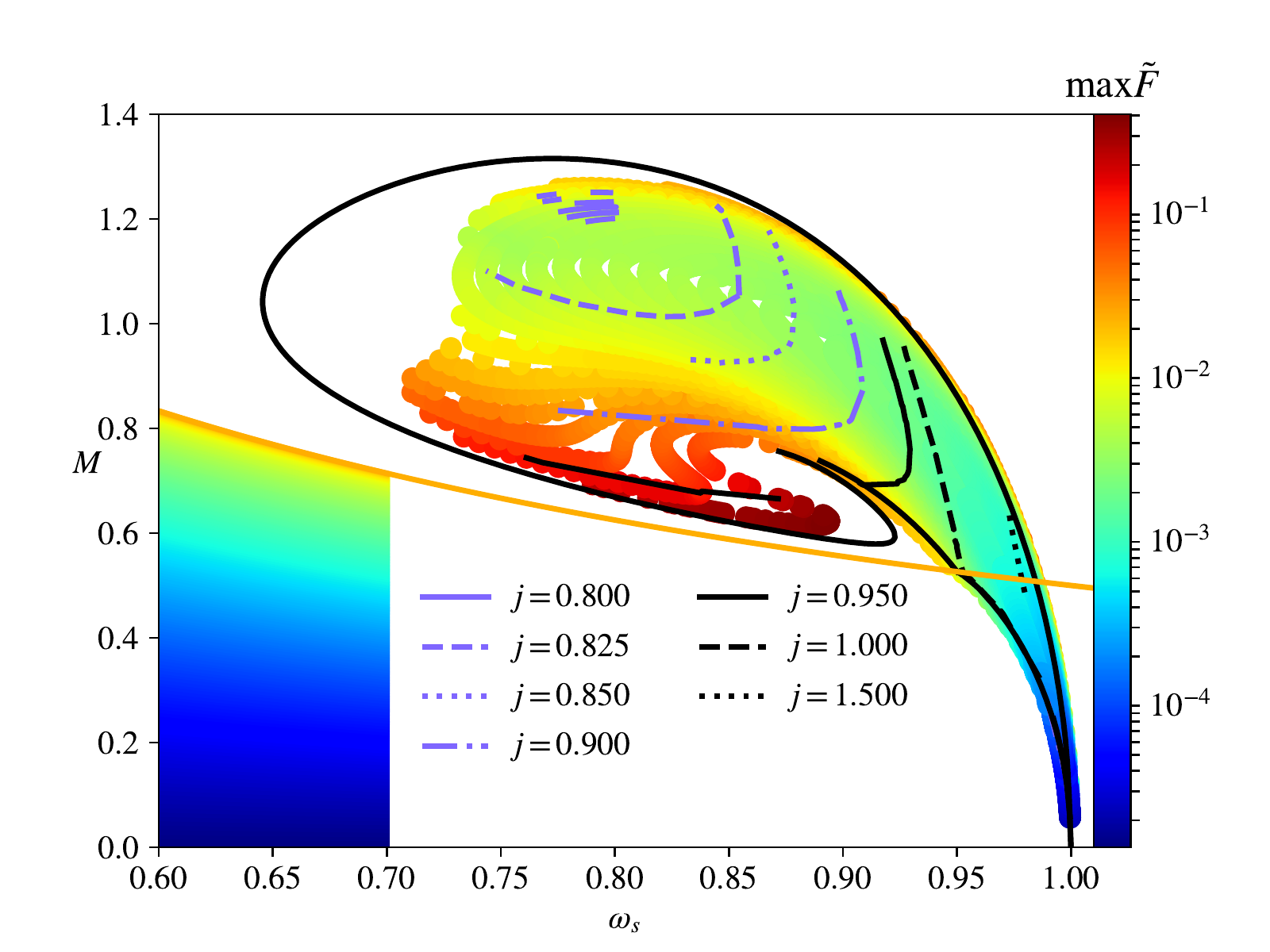}
\caption{Colorplot for the maximum of the normalized flux in the mass $vs$. $\omega_s$ diagram. Some solutions of specific spin parameter are highlighted. On the bottom left corner a sample for Kerr black holes is displayed.}
\label{mo_re}
\end{figure}

The non-uniqueness of the scalarized solutions for a pair of mass and angular momentum is reflected on the profile of the radiant energy flux for a fixed value of $j$. In what follows, we present several cases specified by a certain spin parameter, and plot within the flux for black holes of different normalized charges. Although it is not possible to find a general pattern, some particular traits appear in each set. Near the marginally bound clouds, the maximum of the flux decreases with increasing charge, while the opposite is seen in regions far from the clouds in the $M$ vs. $\omega_s$ diagram. The solutions span a large interval of $j$, but a more prominent set of qualitatively different solutions concentrates for values slightly lower than the Kerr bound $j=1$.  In Fig. \ref{mo_re} we offer a general view of the maximum value of the normalized flux in the same mass vs. $\omega_s$ diagram we displayed before. Some curves of fixed $j$ are shown for illustration purposes, and we note that for any value $j\leq 1$ there will be a small sample of solutions starting at the clouds which are not depicted because they require much higher resolution. Clearly, their sequences are not connected to their respective (same $j$) sets displayed in different regions above the extremal Kerr line.
 The exception happens for the particular case of $j=1$ that forms two curves connected at the three points intersection, namely where the lines of extremal Kerr, extremal KBHsSH and clouds meet. Another interesting case shown here is that of $j=0.8$, where far from the clouds the solutions lie somewhat diffusely on a certain region rather than forming a curve. Thus, for a chosen value of the spin parameter, there might be solutions severely different from one another, which can result in radiant energy fluxes of different orders of magnitude. On the lower left corner of the diagram, the normalized flux for Kerr BHs is shown and as it is known, it increases monotonically with the spin parameter, e.g. see eq. (\ref{masskerr}). From the figure, we observe that the maximum normalized flux for KBHsSH can be an order of magnitude higher than that of extremal Kerr BHs.

\begin{figure}[h]
\includegraphics[width=0.6\textwidth]{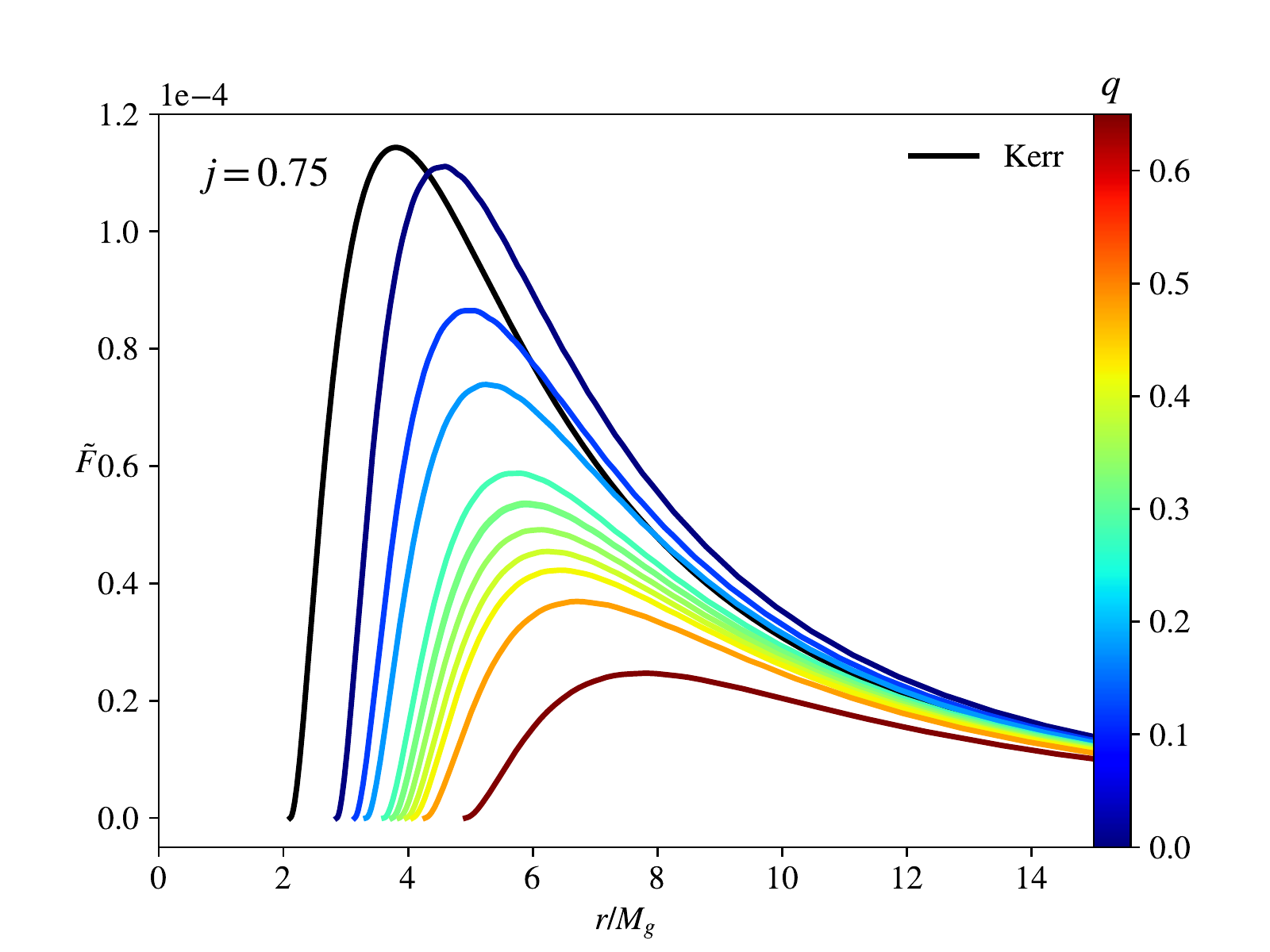}
\caption{Radiant energy fluxes for $j=0.75$ solutions. The color scheme indicates the value of $q$ and therefore how hairy a solution is. }
\label{re075}
\end{figure}

The set of solutions starting at the clouds display a profile with decreasing maximum of the flux for increasing charge, and in this case the bald Kerr BHs feature the stronges radiant energy flux. Sets of lower values of the spin parameter exist only in this region, and an example with $j=0.75$ is given in Fig. \ref{re075}, where as the charge grows larger the profile flattens out.

\begin{figure}
\begin{subfigure}{.5\textwidth}
  \centering
  \includegraphics[width=1.\linewidth]{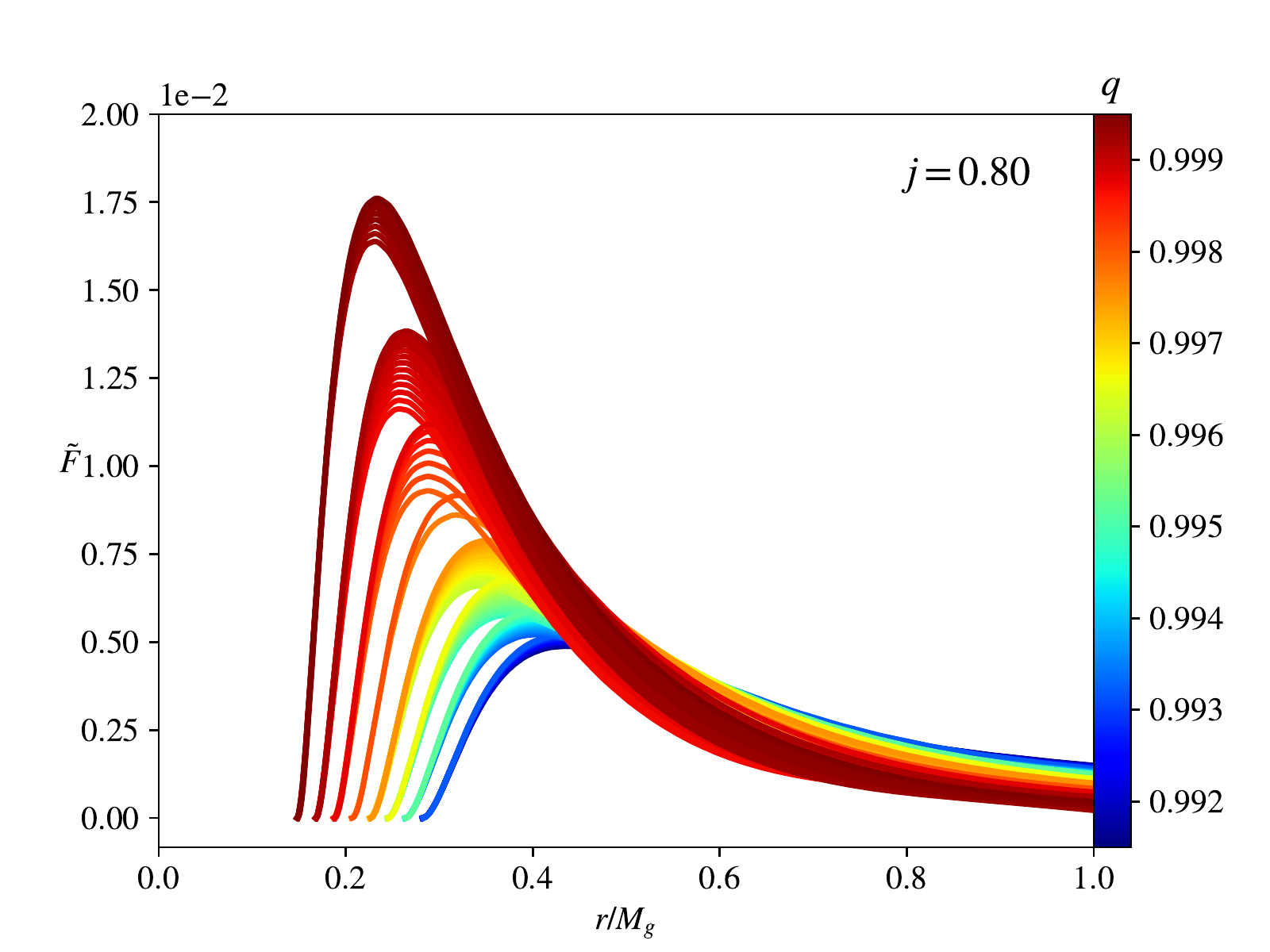}
\end{subfigure}%
\begin{subfigure}{.5\textwidth}
  \centering
  \includegraphics[width=1.\linewidth]{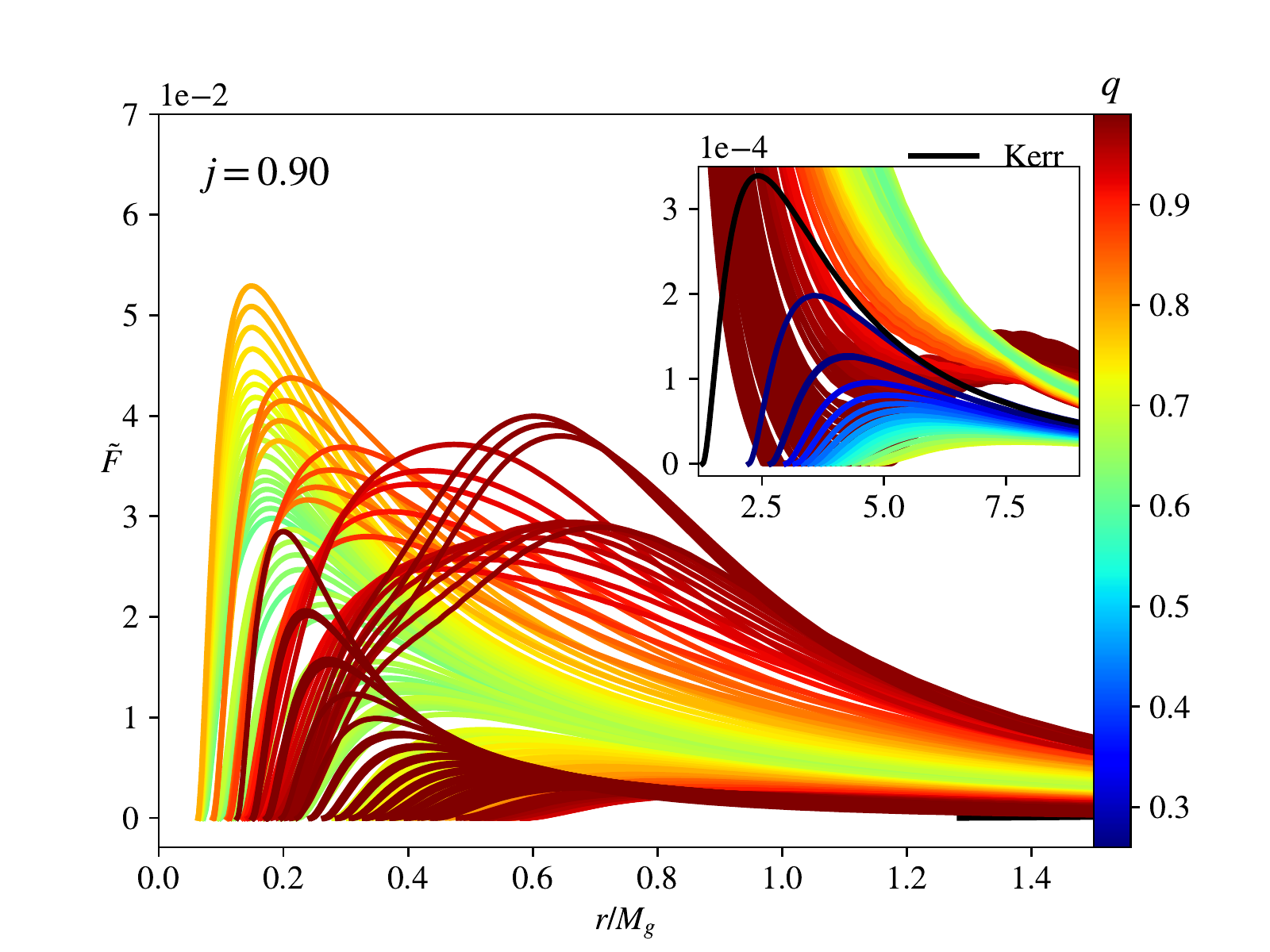}
\end{subfigure}
\\
\begin{subfigure}{.5\textwidth}
  \centering
  \includegraphics[width=1.\linewidth]{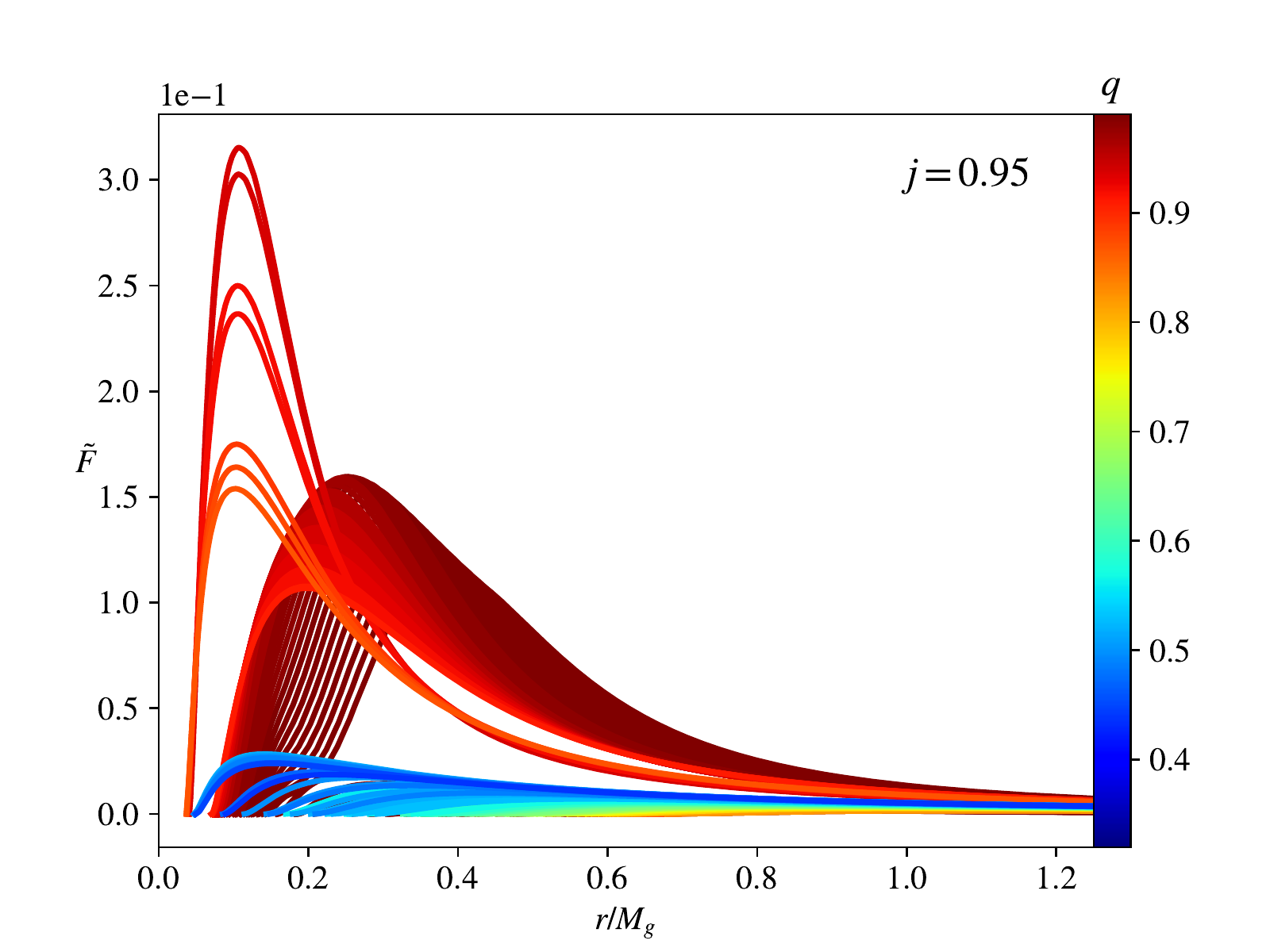}
\end{subfigure}%
\begin{subfigure}{.5\textwidth}
  \centering
  \includegraphics[width=1.\linewidth]{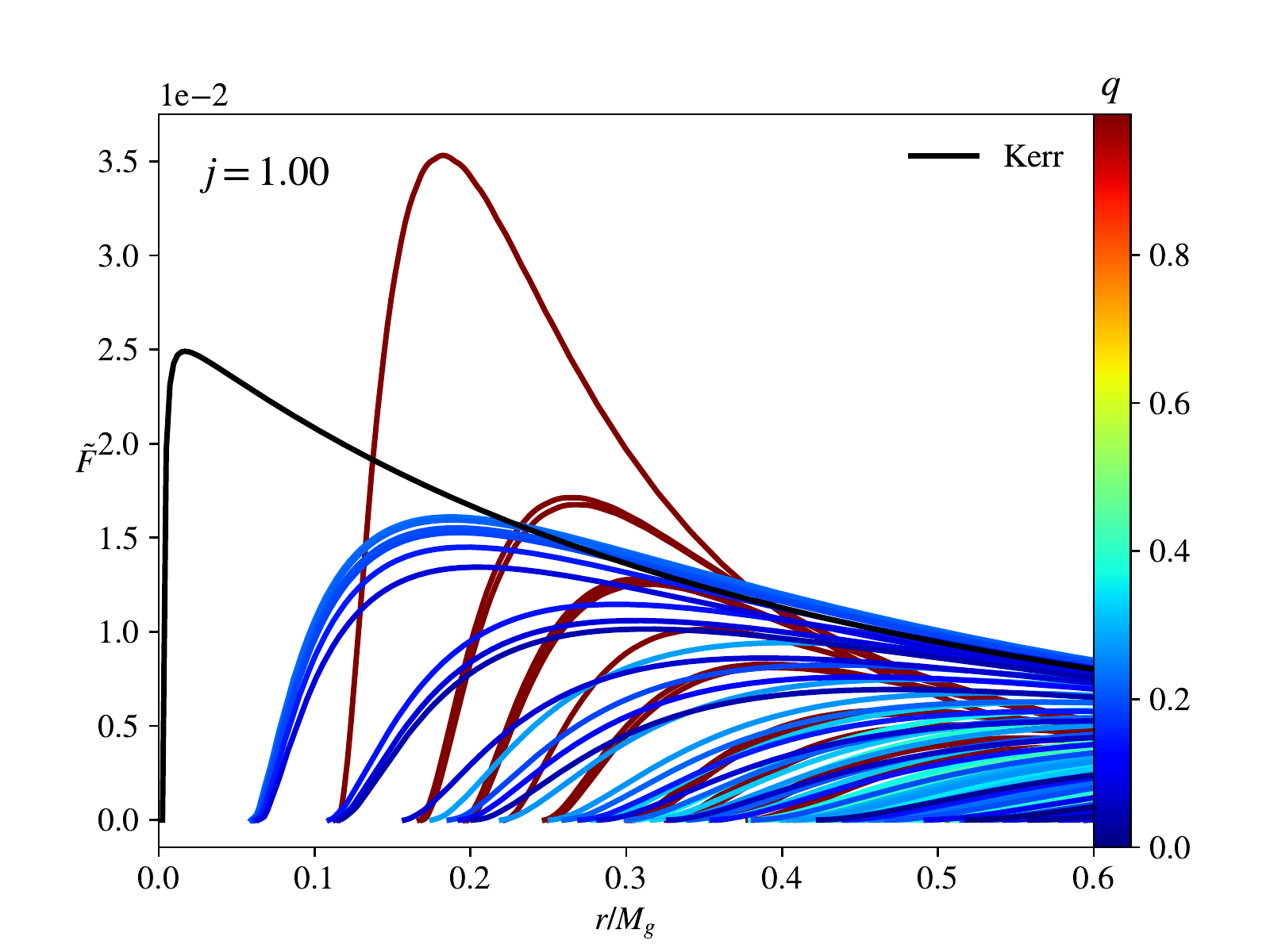}
\end{subfigure}
\\
\begin{subfigure}{.5\textwidth}
  \centering
  \includegraphics[width=1.\linewidth]{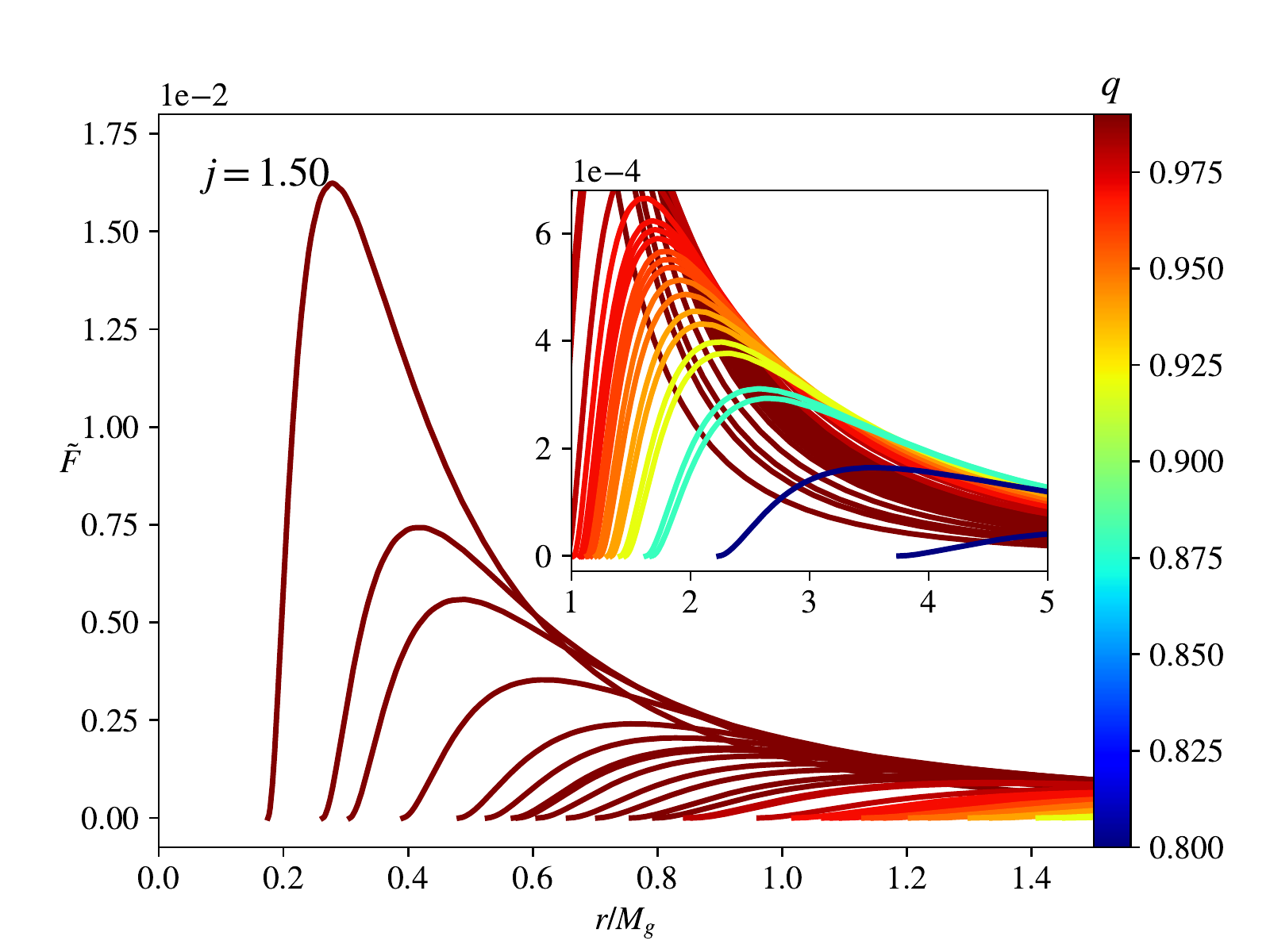}
\end{subfigure}%
\begin{subfigure}{.5\textwidth}
  \centering
  \includegraphics[width=1.\linewidth]{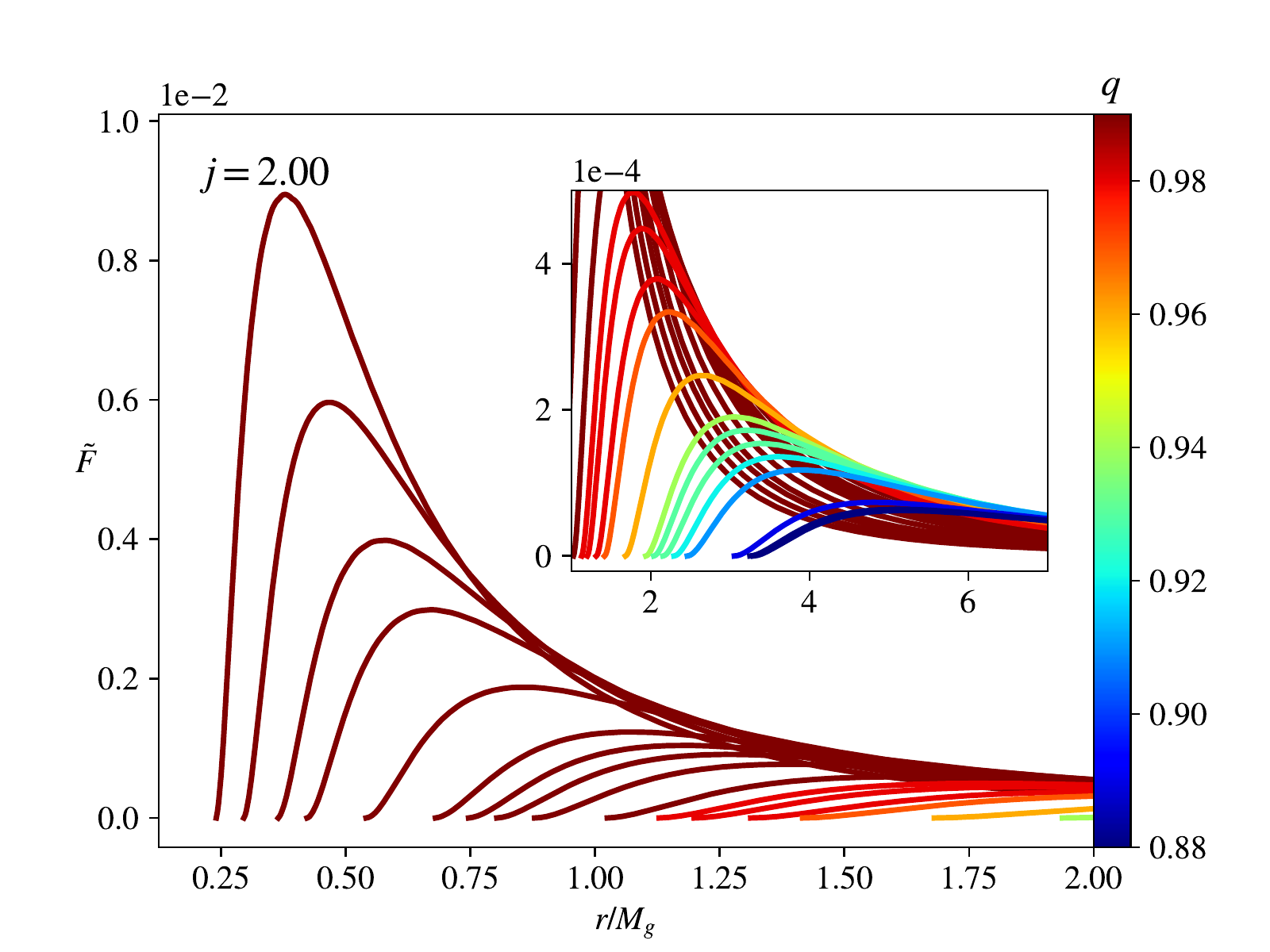}
\end{subfigure}
\caption{Radiant energy fluxes for solutions of fixed $j$.}
\label{re_several}
\end{figure}

This behavior changes in other regions of the parameter space, as can be seen from Fig.\ref{re_several} where six different examples are given. The top left panel displays the flux for solutions featuring $j=0.8$ which are far from the clouds. As mentioned above, this is a special case where the set does not form a curve crossing a large area of the parameter space in Fig. \ref{mo_re} but is concentrated in a more limited region where the solutions are quite similar. Hence, the single behavior observed in the flux profile as the normalized charge varies, i.e. the larger it is, the higher the peak of the flux gets. The top right and middle left panels show solutions for which $j=0.90$ and $0.95$, respectively. Here we enter the regime where a wider region of the parameter space is crossed by a curve of fixed $j$. As we notice from the pictures, there is not a unique way on how the flux profile scales with the scalar charge. On the panel for $j=0.90$ we display an inset to show the flux corresponding to Kerr, along with those solutions close to the cloud for which the maximum of the flux decreases with increasing charge. More importantly, we notice that the highest flux peak found is over two orders of magnitude greater than the Kerr peak.
The middle right panel depicts solutions for which $j=1$, the quite special case since here the curve lying near the clouds is connected to the one existing above the extremal Kerr line. Furthermore, we observe that the fluxes are of the same order of magnitude. The bifurcation from the extremal Kerr is made clear: for the solutions lying above this line, the peak of the flux increases with the scalar charge, whereas for those below it, the peak decreases with increasing charge. Nevertheless, we remark that interestingly the radius of the ISCO increases in both cases. Finally, in the bottom panel we show fluxes for solutions with a spin parameter above the Kerr bound, which never approach the clouds but remain mainly in a narrow region of the parameter space for high values of $\omega_s$. In these cases the peak increases with the charge, and we note that albeit hard to establish numerically: the higher $j>1$ is, the larger is the lowest charge found in the solution set.

\subsection{Efficiency}

\begin{figure}[h]
\includegraphics[width=0.6\textwidth]{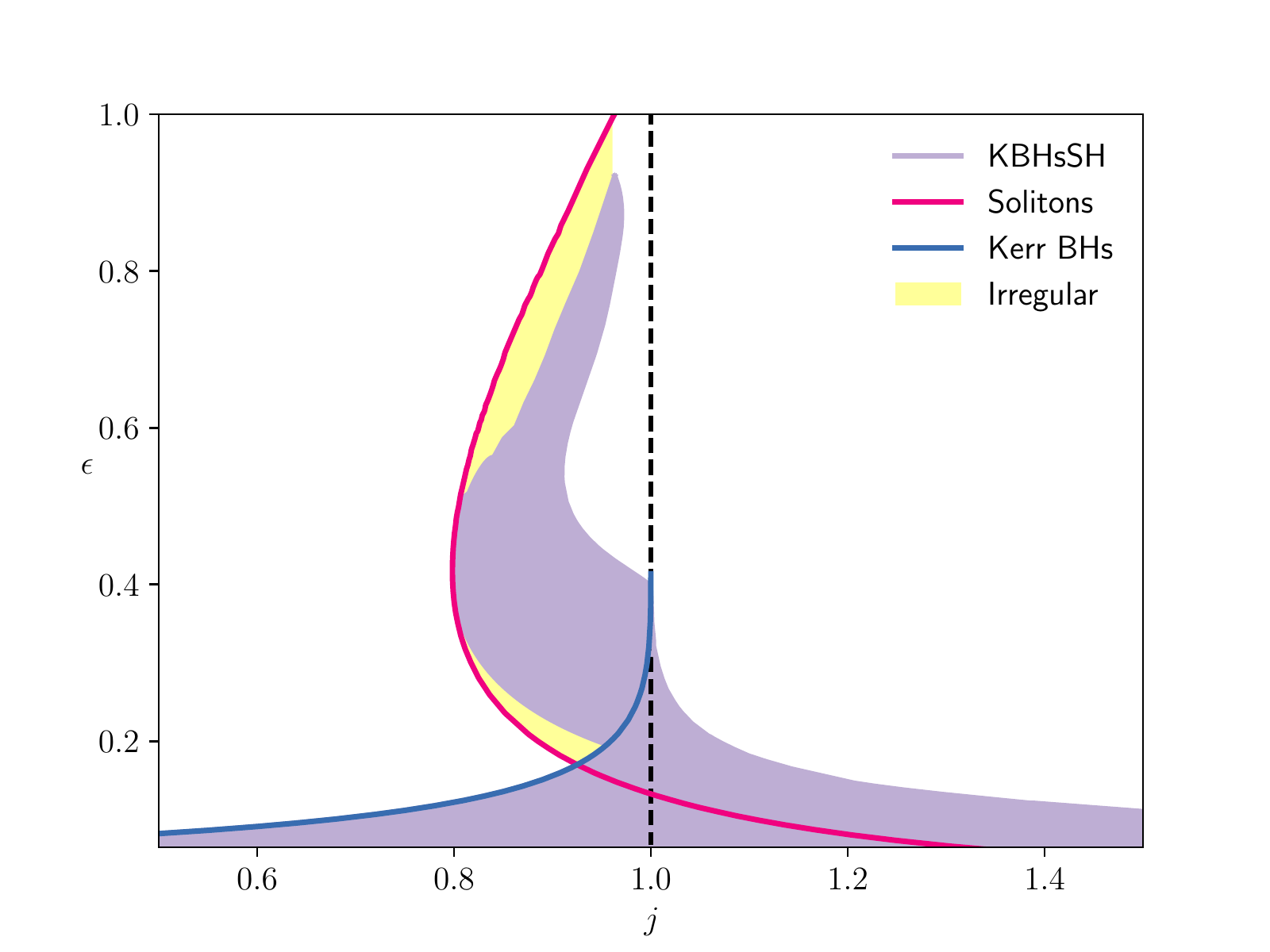}
\caption{Efficiency $vs$. $j$. Note that there is no uniqueness in this domain, i.e. for a pair $(j,\epsilon)$.   }
\label{eff}
\end{figure}

Black holes offer the most efficient mechanism for converting rest energy into radiation. As stated earlier, this parameter is given by the difference between the particle's normalized energy at infinity - which is one - and the energy at the ISCO, see i.e. eq. (\ref{eqeff}). Fig. \ref{eff} offers the efficiency of KBHsSH against $j$, where the limiting cases of Kerr BHs and solitons are highlighted in two different curves. Heeding the peculiarities discussed in Section \ref{sec:metricisco}, we consider only \emph{regular} solutions, lying outside the shaded region of Fig. \ref{domainskbh}. This parameter space is not unique and several solutions can be found for a given pair of ($j$, $\epsilon$).

Given a specific value of $j<1$, KBHsSH can always be found below the Kerr curve. At the Kerr bound (and above), all solutions are less efficient than the bald Kerr, in agreement with Fig \ref{re_several}, for the radius of the ISCO increases with the scalar charge. However, for the interval $j\in (0.8,1.0)$ we find a range in the diagram well above the Kerr line, and the highest efficient found exceeds $90\%$. This is only possible for black holes with very small horizon radius and very large scalar charge as they behave more strongly as solitons than black holes and the ISCO is found at very small radii.

\subsection{Luminosity}

\begin{figure}
\begin{subfigure}{.5\textwidth}
  \centering
  \includegraphics[width=1.\linewidth]{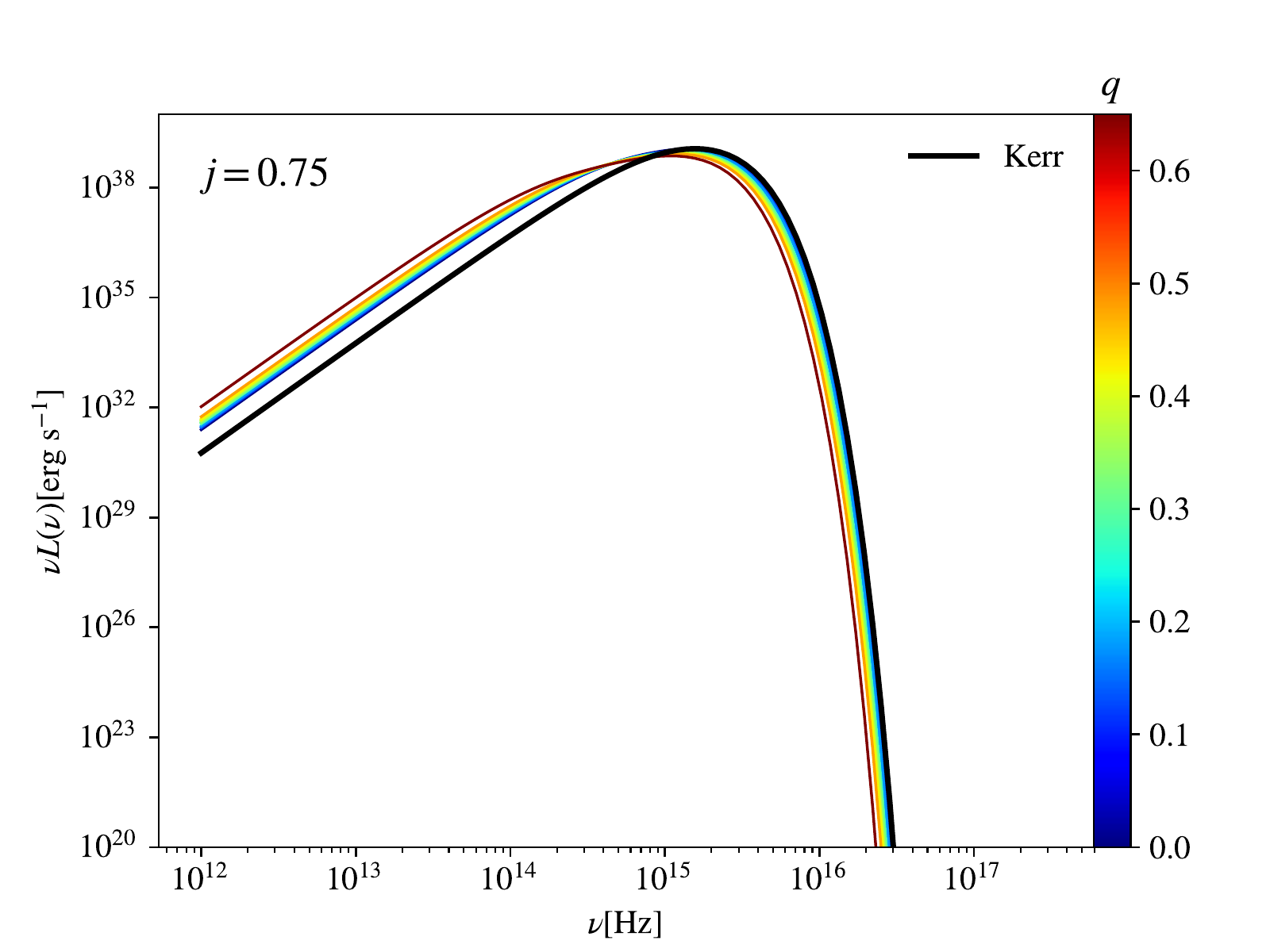}
\end{subfigure}%
\begin{subfigure}{.5\textwidth}
  \centering
  \includegraphics[width=1.\linewidth]{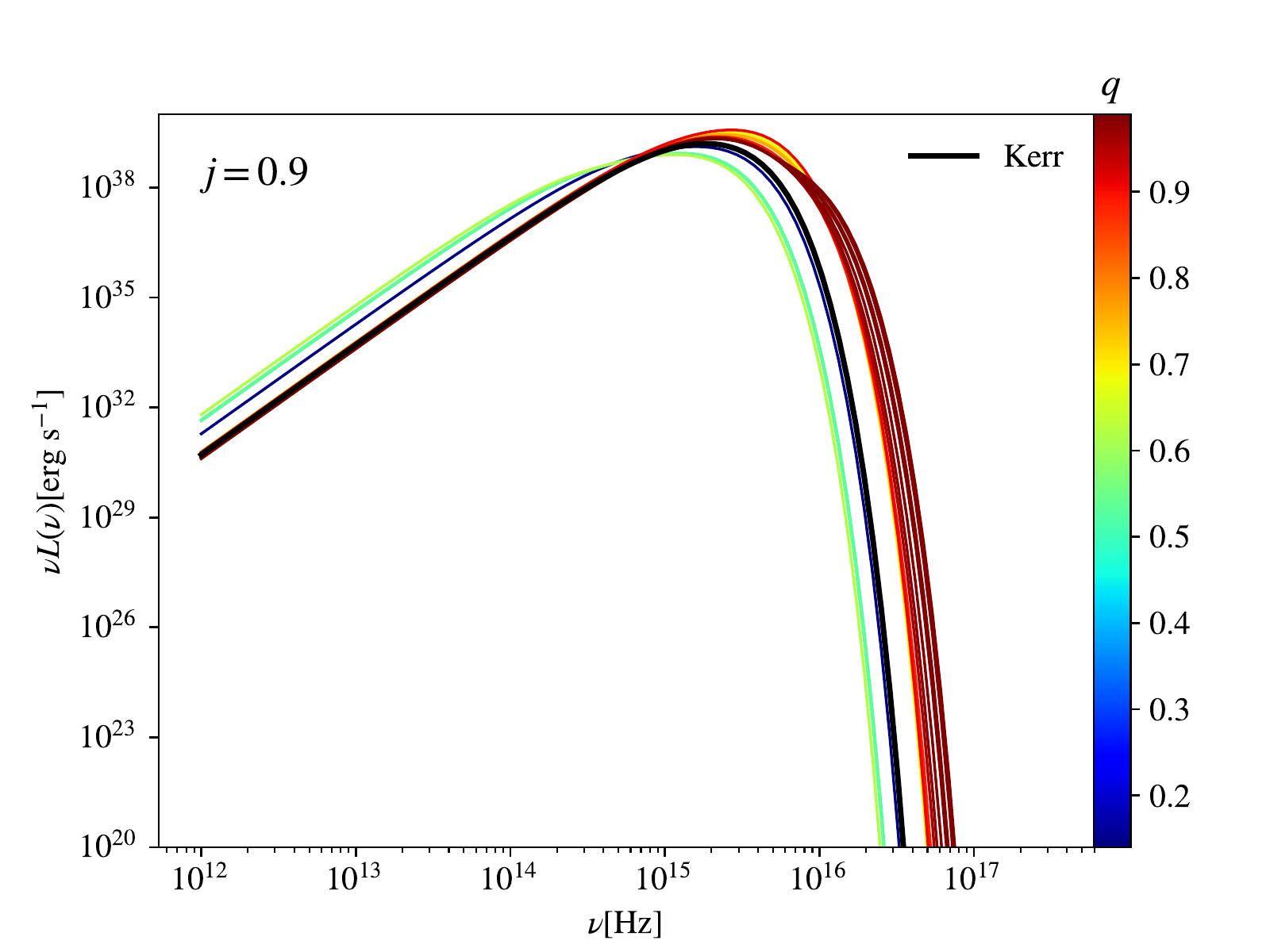}
\end{subfigure}
\\
\begin{subfigure}{.5\textwidth}
  \centering
  \includegraphics[width=1.\linewidth]{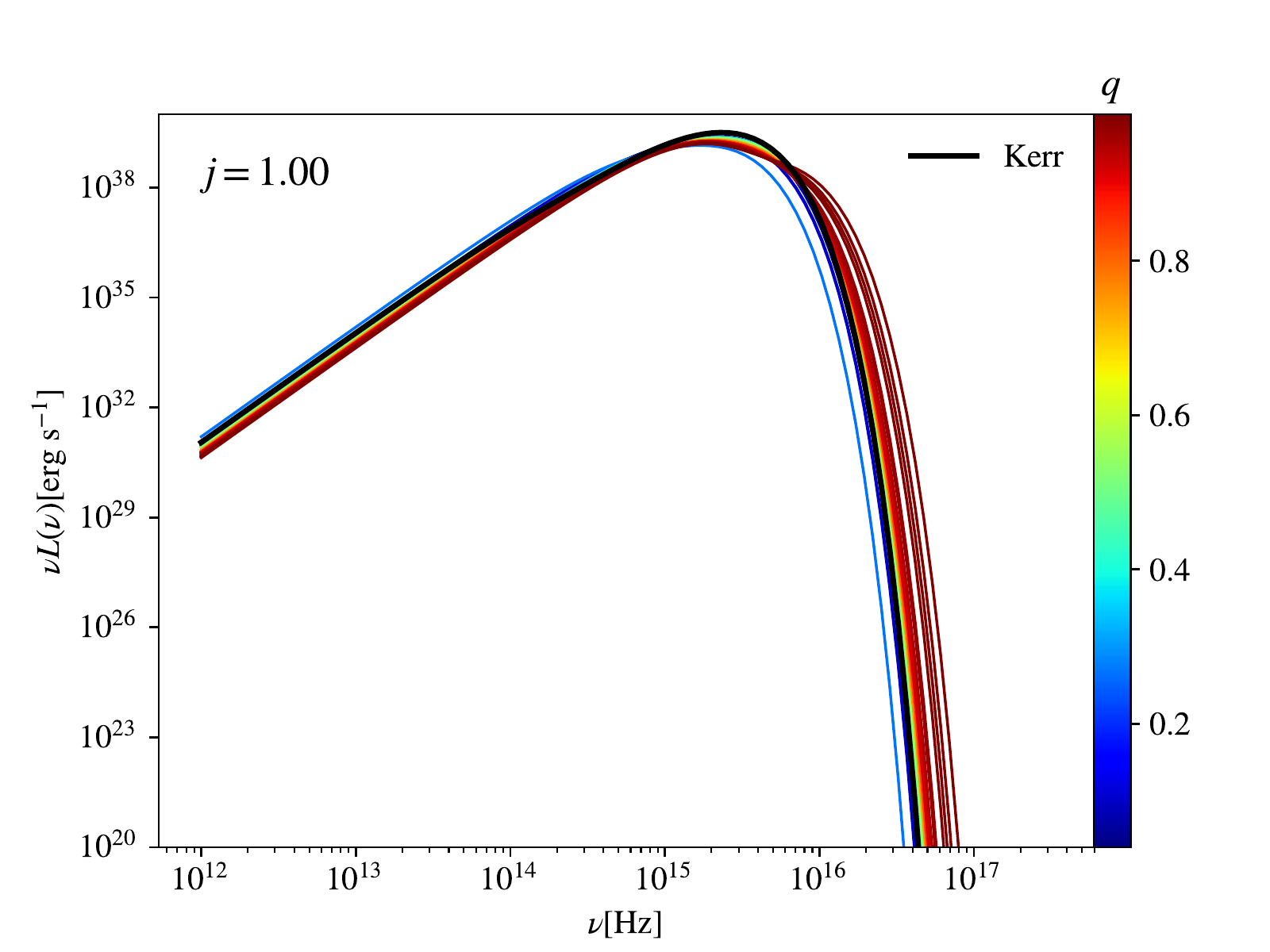}
\end{subfigure}%
\begin{subfigure}{.5\textwidth}
  \centering
  \includegraphics[width=1.\linewidth]{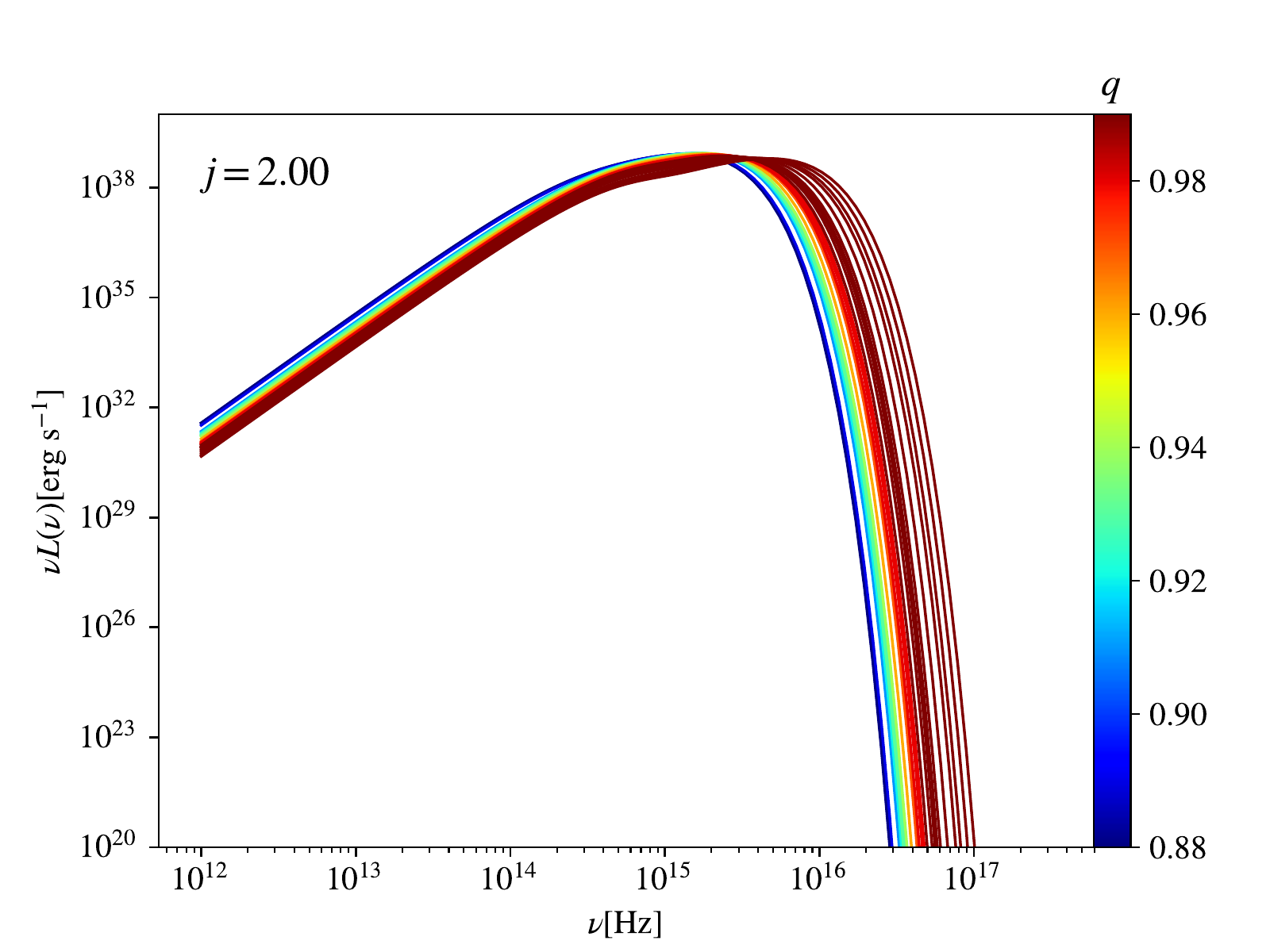}
\end{subfigure}
\caption{Luminosities for solutions of fixed $j$.}
\label{fig:lum}
\end{figure}

Let us now evaluate the observed luminosity for a few scalarized black holes. We assume the central object to be a supermassive black hole of $M=2.5\times 10^6 M_\odot$ and $\dot{M}=2.0\times 10^{-6}M_\odot/$yr, with the plane of the disk oriented at zero degrees with the euclidean line path to the observer. These values are chosen for illustration purposes only, but we stress that they are fairly conservative (much below the Eddington limit) and reasonable to represent the supermassive BHs. We also consider four different cases of fixed spin parameter separately, i.e. $j\in\{0.75, 0.9, 1.0, 2.0\}$. The results are displayed in Fig. \ref{fig:lum}, together with the curves corresponding to the respective Kerr BH (when applied). For the chosen mass scale and accretion rate, the luminosity peaks at the visible-UV interface part of the electromagnetic spectrum, while stellar mass BHs with similar mass-accretion rate ratio $\dot{M}/M$ would lie on the X-ray band. In all cases below the Kerr bound, some solutions naturally show negligible differences with the bald BH case (also part of the solution set). As we get further away from the clouds in the parameter space, though, observable discrepancies arise.

The particular case of $j=0.75$ has a peak in luminosity for $q=0.0$ (Kerr BHs). Recall that the efficiency for this spin parameter decreases with the charge, and so does the radiant energy flux. The luminosity curves intersect each other, increasing and decreasing faster for larger charges as the  frequency increases. Therefore, at lower frequencies $\nu L$ is higher for larger charges and at higher frequencies $\nu L$ is higher for lower charges. The largest difference in the peak of luminosity between the curves is of $36,7\%$ whereas for the peaking frequency it is of $31.6\%$.

As previously shown, there are many qualitatively different solutions with fixed spin parameters in the interval $j\in (0.8,1.0)$. This is apparent in the luminosity profile as curves can fall either below of above Kerr case. For $j=0.9$ in the examples depicted in the figure, the peak of luminosity is highest for $q=0.91$ and lowest for $q=0.55$, with a relative difference of $77.2\%$, and of $53.3\%$ in the peaking frequency. Extremal Kerr displays the highest luminosity peak among the curves of $j=1.0$, but interestingly not the highest peaking frequency, which is offset by $14.4\%$. The largest relative difference in the peak of luminosity is of $49.4\%$. Above the Kerr bound most solutions are highly scalarized and much closer to a solitonic star than to a bald black hole. For $j=0.2$, the peak of luminosity is proportional to the charge, while peaking frequency decreases with it. In all cases, the differences are much higher at the end of the observed spectrum, and the interval for the cutoff frequencies (where $\nu L=0$) grows consistently with $j$, with an interval of $1.25\times 10^{16}$Hz for $j=0.75$ and of $1.5\times 10^{17}$Hz for $j=2.0$.

\section{Conclusions}
\label{sec:conclusions}
In the present article we analyzed some features of circular orbit geodesics around KBHsSH, and calculated the radiant energy flux and luminosity for several solutions through the thin accretion disk approach. We showed that due to the non-monotonic behavior of the metric function on the equatorial plane some peculiar traits appear for a subset of the solutions, and categorized them accordingly. Specifically, some solutions feature the static ring, where one of the eigenvalues for the orbital velocity is null, some contain regions where both eigenvalues are positive and the orbit is stable, some contain disconnected regions of stable orbits for the eigenvalue which is everywhere positive, and finally some for which there is a region where no real valued eigenvalues exist and hence no inertial circular orbits are possible. It is unclear how accretion would take place in those backgrounds, and the thin disk approach is certainly not suitable for most of them, where discontinuities appear for both the orbital velocity and the angular momentum. 

The regular solutions, i.e. the set that does not suffer from such discontinuities, were normalized with respect to their masses and analyzed in bundles of same spin parameter, but different charges. Wherever $j\leq 1.0$, we also computed the quantities at matter for the corresponding Kerr BH. Concerning the flux and luminosity, because of the non-uniqueness of the solutions for any chosen $j$, many different profiles appear, and although we observe some pattern in their dependence on the scalar charge for smaller or higher $j$, in the interval $j\in(0.8,1.0)$ wide differences appear even for solutions of similar $q$. The ratio between the largest and smallest peak of the normalized flux is about $\tilde{F}_{\max}/\tilde{F}_{\min}\sim 10^4$. 

The efficiency of KBHsSH is much varied, and always below that of Kerr with the same spin parameter for $j<0.8$, and below that of extremal Kerr for $j>1.0$. In the interval in between, however, it takes various values, reaching over $90\%$ for some of the most hairy solutions where the ISCO radius approaches that of the horizon which is already quite small.

The luminosity profiles do not differ strongly enough to peak at different bands of the spectrum or to be over many orders of magnitude apart, but the differences are still significant to be measured. While many results depart very slightly from what is obtained from a Kerr BH (which is part of the solution set), others can peak at lower or higher frequencies, as well as be more or less luminous for some fixed $\nu_e$ and $j$.

Even though Kerr models fit the data well, the existence of an extra parameter $q$ describing KBHsSH offers a manifold extension in the solution set which makes it hard to verify, constrain or dismiss their existence solely based on this kind of observations. On the other hand, once the mass, accretion rate and angular momentum are well constrained, it is possible to infer from accretion disks observations what the scalar charge of the BH is. Such indirect measurement is fundamental in order to specify a particular compact object since $Q$ cannot be extracted asymptotically from the metric as there is no Gauss law associated with it. By combining the luminosity with other 
observations such as from Iron K$\alpha$ line and shadows, one could strongly narrow down the possible solutions with a good enough fit, and assess how KBHsSH perform in contrast to bald Kerr. More stringent constraints are definitely to be expected from future measurements of geodesic motion of nearby stars. As we have shown with the peculiar structure of circular orbits on the equatorial plane for a subset of solutions, the growth of a massive hair with off-center energy distribution can cause the metric functions to behave non-monotonically, as opposed to what happens for GR BHs outside the event horizon. 

\section*{Acknowledgements}
LC and DD acknowledge financial support via an Emmy Noether Research Group funded by the German Research Foundation (DFG) under grant no. DO 1771/1-1. SY would like to thank the University of Tuebingen for the financial support.  SY acknowledges financial support by the Bulgarian NSF Grant KP-06-H28/7. Networking support by the COST Actions  CA16104 and CA16214 is also gratefully acknowledged.

\appendix
\section{Radiant Energy Flux in Adapted Spherical Coordinates}
\label{app:refsc}
The equations describing thin accretions disks widely used in the literature were first derived in \cite{Novikov:1973kta,Page:1974he}, where the authors employed adapted cylindrical coordinates \emph{near the equatorial plane}. Some of these equations are revisited here for a metric written over adapted spherical coordinates since there are a few subtleties one needs to heed. The end result, though, is exactly the same in this approximation regime.

As in the original paper, quantities written inside brackets are averaged over time and the axial coordinate:
\be
\langle\Psi(r,\theta)\rangle\equiv\frac{1}{2\pi\Delta t}\int_0^{\Delta t}\int_0^{2\pi}\Psi(t,r,\theta,\varphi) d\varphi dt.
\ee

The four-velocity of the fluid in its local rest frame, $u^\mu_\text{inst}$, is averaged also over the height of the disc and weighted by its rest mass density $\rho_0$,
\be
u^\mu\equiv\frac{1}{\Sigma}\int_{\pi-\theta_0}^{\theta_0}\langle\rho_0u^\mu_\text{inst}\rangle\sqrt{g_{\theta\theta}}d\theta,
\ee
where $\Sigma$ is the averaged surface energy density at a certain point
\be
\Sigma(r)\equiv\int_{\pi-\theta_0}^{\theta_0}\langle\rho_0\rangle\sqrt{g_{\theta\theta}}d\theta.
\ee

The energy-momentum tensor of the fluid is written in terms of the averaged four-velocity in general form as
\be
T^{\mu\nu}=\rho_0u^\mu u^\nu+t^{\mu\nu}+2u^{(\mu}q^{\nu)},
\ee
where $t^{\mu\nu}$ is the averaged energy-momentum tensor in the rest frame and is orthogonal to $u^\mu$, and $q^\mu$ is a energy flux vector field, also orthogonal to the four-velocity.

Among all of the assumptions made and used to derive the equations is the assertion that the disc is optically thick, i.e. the radiation emitted along the disc's plane is negligible, and one considers only the flux orthogonal to its face (hence $q^\mu u_\mu=0$). Thus, the averaged flux of radiant energy is given by
\be
F(r)\equiv\langle q^\theta(r,\theta_0)\sqrt{g_{\theta\theta(r,\theta_0)}}\rangle=\langle q^\theta(r,\pi-\theta_0)\sqrt{g_{\theta\theta(r,\pi-\theta_0)}}\rangle,
\ee
which is simply the tensor transformation from the original vertical flux $q^z$ when $z\ll r$ or $\theta_0\approx\pi/2$. Finally, the explicit form of the flux is derived from the conservation laws, yielding
\be
F(r) = -\frac{c^2\dot{M}}{4\pi\sqrt{-g/g_{\theta\theta}}}\frac{\partial_r\Omega}{(E-\Omega L)^2}\int_{r_{ISCO}}^r (E-\Omega L)\partial_rLdr.
\ee
\bibliographystyle{ieeetr}
\bibliography{biblio}
\end{document}